\documentclass[11pt]{article}
\usepackage{graphicx}
\usepackage{amssymb}
\usepackage{amsmath}
\usepackage{epsfig}

\def\cadremath#1{\vbox{\hrule\hbox{\vrule\kern8pt\vbox{\kern8pt
			\hbox{ {$\displaystyle #1 $ } }\kern8pt} 
			\kern8pt\vrule}\hrule}}

\def\today{\number\day\space\ifcase\month\or Janvier \or F\'evrier \or  Mars
   \or Avril \or Mai \or Juin \or Juillet \or Ao\^ut \or Septembre \or Octobre
   \or Novembre \or D\'ecembre \fi\number \year}

\title{A semiclassical study of the Jaynes-Cummings model.}
\author{O. Babelon,  L. Cantini, B. Dou\c{c}ot.}
\date{8 May 2009}                                           % Activate to display a given date or no date

\begin{document}
\maketitle
\centerline{Laboratoire de Physique Th\'eorique et Hautes Energies\footnote{ 
LPTHE, Tour 24-25, 5\` eme \'etage, Boite 126, 4 Place Jussieu,  75252 Paris Cedex 05.}}
\centerline{Unit\'e Mixte de Recherche UMR 7589}
\centerline{Universit\'e Pierre et Marie Curie-Paris6; CNRS;}

\bigskip
{\bf Abstract.} We consider the Jaynes-Cummings model of a single quantum spin $s$ coupled
to a harmonic oscillator in a parameter regime where the underlying classical dynamics
exhibits an unstable equilibrium point. This state of the model is relevant to the physics
of cold atom systems, in non-equilibrium situations obtained by fast sweeping through a
Feshbach resonance. We show that in this integrable system with two degrees of freedom,
for any initial condition close to the unstable point, the classical dynamics is
controlled by a singularity of the focus-focus type. In particular, it displays the expected
monodromy, which forbids the existence of global action-angle coordinates.  Explicit
calculations of the joint spectrum of conserved quantities reveal the monodromy at the
quantum level, as a dislocation in the lattice of eigenvalues. We perform a detailed semi-classical
analysis of the associated eigenstates. Whereas most of the levels are well described by the usual
Bohr-Sommerfeld quantization rules, properly adapted to polar coordinates, we show how these
rules are modified in the vicinity of the critical level. The spectral decomposition of the
classically unstable state is computed, and is found to be dominated by the critical WKB states.
This provides a useful tool to analyze the quantum dynamics starting from this particular state,
which exhibits an aperiodic sequence of solitonic pulses with a rather well defined characteristic 
frequency.

\section{Introduction.}

The Jaynes-Cummings model was originally introduced to describe the near resonant
interaction between a two-level atom and a quantized mode of the electromagnetic field~\cite{JC}.
When the field is treated classically, the populations of the two levels exhibit
periodic Rabi oscillations whose frequency is proportional to the field intensity.
The full quantum treatment shows that the possible oscillation frequencies are 
quantized, and are determined by the total photon number stored in the mode. For an
initial coherent state of the field, residual quantum fluctuations in the
photon number lead to a gradual blurring and a subsequent collapse of the Rabi oscillations
after a finite time. On even longer time scales, the model predicts a revival of the
oscillations, followed later by a second collapse, and so on. These collapses and revivals
have been analyzed in detail by Narozhny et al~\cite{NSE}, building on the exact solution
of the Heisenberg equations of motion given by Ackerhalt and Rzazewski~\cite{AR}.
Such complex time evolution has been observed experimentally on the Rydberg micromaser~\cite{Rempe87}.
More recently, a direct experimental evidence that the possible oscillation frequencies  
are quantized has been achieved with Rydberg atoms interacting with the small coherent fields
stored in a cavity with a large quality factor~\cite{Brune96a}. Another interesting feature
of the Jaynes-Cummings model is that it provides a way to prepare the field in a linear
superposition of coherent states~\cite{Banacloche91}. In the non resonant case, closely related ideas
were used to measure experimentally the decoherence of Schr\"odinger cat states of the 
field in a cavity~\cite{Brune96b}.

In the present paper, we shall consider the generalization of the Jaynes-Cummings model where
the two-level atom is replaced by a single spin $s$. One motivation for this is the phenomenon of
superradiance, where a population of identical two-level atoms interacts coherently with the
quantized electromagnetic field. As shown by Dicke~\cite{Dicke}, this phenomenon can be viewed
as the result of a cooperative behavior, where individual atomic dipoles build up to make
a macroscopic effective spin. In the large $s$ limit, and for most initial conditions, a semi-classical
approach is quite reliable. However, the corresponding classical Hamiltonian system with two degrees
of freedom is known to exhibit an unstable equilibrium point, for a large region in its parameter space.
As shown by Bonifacio and Preparata~\cite{Bonifacio}, the subsequent evolution of the system, starting
from such a state, is dominated by quantum fluctuations as it would be for a quantum pendulum
initially prepared with the highest possible potential energy. These authors have found that, at short times,
the evolution of the system is almost periodic, with solitonic pulses of photons separating quieter
time intervals where most of the energy is stored in the macroscopic spin. Because the stationary states
in the quantum system have eigenenergies that are not strictly equidistant, this quasiperiodic behavior
gives way, at longer times, to a rather complicated pattern, that is reminiscent of the collapses and
revivals in the $S=1/2$ Jaynes-Cummings model.

An additional motivation for studying the large $s$ Jaynes-Cummings model comes from recent
developments in cold atom physics. It has been shown that the sign and the strength of the
two body interaction can be tuned at will in the vicinity of a Feshbach resonance, and this
has enabled various groups to explore the whole crossover from the Bose Einstein condensation
of tightly bound molecules~\cite{Greiner03} to the BCS condensate of weakly bound atomic 
pairs~\cite{Bourdel04,Zwierlein04}. In the case of a fast sweeping of the external magnetic field
through a Feshbach resonance, some coherent macroscopic oscillations in the molecular condensate
population have been predicted theoretically~\cite{Levitov}, from a description of the low
energy dynamics in terms of a collection of $N$ spins $1/2$ coupled to a single harmonic oscillator.
This model has been shown to be integrable~\cite{YKA} by Yuzbashyan et al. who emphasized its connection
with the original integrable Gaudin model~\cite{Gaudin83}. It turns out that in the cross-over region,
the free Fermi sea is unstable towards the formation of a pair condensate, and that this instability
is manifested by the appearance of two pairs of conjugated complex frequencies in a linear analysis.
This shows that for any value of $N$ (which is also half of the total number of atoms), these coherent
oscillations are well captured by an effective model with two degrees of freedom, one spin, and one
oscillator. Therefore, there is a close connection between the quantum dynamics in the neighborhood
of the classical unstable point of the Jaynes-Cummings model, and the evolution of a cold Fermi gas after
an attractive interaction has been switched on suddenly.    

The problem of quantizing a classical system in the vicinity of an unstable
equilibrium point has been a subject of recent interest, specially in the mathematical
community. The Bohr-Sommerfeld quantization principle suggests that the density of states
exhibits, in the $\hbar$ going to zero limit, some singularity at the energy of the
classical equilibrium point, in close analogy to the Van Hove singularities for the
energy spectrum of a quantum particle moving in a periodic potential. This intuitive
expectation has been confirmed by rigorous~\cite{Colin94a,Colin94b,Brummelhuis95}
and numerical~\cite{Child98} studies. In particular, for a critical level of a system
with one degree of freedom, there are typically $|\ln \hbar|$ eigenvalues in an energy interval
of width proportional to $\hbar$ around the critical value~\cite{Colin94b,Brummelhuis95}.
A phase-space analysis of the corresponding wave-functions shows that they are concentrated
along the classical unstable orbits which leave the critical point in the remote past and
return to it in the remote future~\cite{Colin94a}. To find the eigenstates requires
an extension of the usual Bohr-Sommerfeld rules, because matching the components of the
wave-function which propagate towards the critical point or away from it is a special
fully quantum problem, which can be solved by reduction to a normal form~\cite{Colin94a,Colin99}.

Finally, we will show that the spin $s$ Jaynes-Cummings model is an example
of an integrable
system for which it is impossible to define global action-angle coordinates.
By the Arnold-Liouville theorem, classical
integrability implies that phase space is foliated by $n$-dimensional
invariant tori. Angle coordinates are
introduced by constructing $n$-independent periodic Hamiltonian flows on
these tori. Hence each invariant torus is equipped with a lattice of  symplectic
translations which act as the identity on this torus, the lattice of periods
of these flows, which is equivalent to the data of $n$ independent cycles on the torus. 
To get the angles we still must choose an origin on these cycles.
All this can be done in a continuous way for close enough nearby tori
showing the existence of local action angle variables.
Globally, a first obstruction can come from the impossibility of choosing an
origin on each torus in a consistent way. In the case of the Jaynes-Cummings
model, this obstruction is absent (see \cite{Duistermat80}, page 702). The
second obstruction to the existence of global action-angle variables comes from the
impossibility of choosing a basis of cycles on the tori in a uniform way.
More precisely, along each curve in the manifold
of regular invariant tori, the lattice associated to each torus can be
followed by continuity.
This adiabatic process, when carried along a closed loop, induces an
automorphism on the lattice attached to initial (and final) torus, 
which is called the monodromy ~\cite{Duistermat80}. 
Several simple dynamical systems, including the spherical
pendulum~\cite{Duistermat80} or the the champagne bottle potential~\cite{Bates91} have been
shown to exhibit such phenomenon. After quantization, classical monodromy induces topological
defects such as dislocations in the lattice of common eigenvalues of the mutually commuting
conserved operators~\cite{San99}. Interesting applications have been found, specially
in molecular physics~\cite{Sadovskii06}. In the Jaynes-Cummings model, the monodromy is
directly associated to the unstable critical point, because it belongs to a singular
invariant manifold, namely a pinched torus. This implies that the set of regular tori is not
simply connected, which allows for a non-trivial monodromy, of the so-called 
focus-focus type~\cite{Zou92,Zung97,Cushman01}. The quantization of a generic system which
such singularity has been studied in detail by V\~u Ng\d{o}c~\cite{San00}. Here, we present
an explicit semi-classical analysis of the common energy and angular momentum eigenstates
in the vicinity of their critical values, which illustrates all the concepts just mentioned.
 
\bigskip

In section~[\ref{classical}] we introduce the classical Jaynes-Cummings model and describe its stationnary points, stable and unstable. We then explain and compute  the classical monodromy when we loop  around the unstable point. We also introduce a reduced system that will be important in subsequent considerations. In section~[\ref{quantique}] we define the quantum model and explain that the appearence of  a default  in the joint spectrum of the two commuting quantities is directly related to the monodromy phenomenon in the classical theory. 
In section~[\ref{semiclassique}] we perform the above reduction directly on the quantum system and study its Bohr-Sommerfeld quantization. The reduction procedure favors some natural coordinates which however introduce some subtleties in the semi classical quantization : there is a subprincipal symbol and moreover the integral of this subprincipal symbol on certain trajectories may have unexpected jumps.We explain this phenomenon. The result is that the Bohr-Sommerfeld quantization works very well everywhere, except for energies around the critical one. In section~[\ref{semiclassiquesing}]  we therefore turn to the semiclassical analysis of the system around the unstable stationnary point. We derive the singular Bohr-Sommerfeld rules. Finally in section~[\ref{evolution}] we apply the previous results to the calculation of the time evolution of the molecule formation rate, starting from the unstable state. For this we need the decomposition of the 
unstable state on the eigenstates basis. We find that only a few states contribute and there is a drastic reduction of the dimensionality of the relevant Hilbert space. To compute these coefficients we remark that it is enough to solve the time dependent Schr\"odinger equation for small time. We perform this analysis and we show that we can extend it by gluing it to the time dependant WKB wave function. This gives new insights on the old result of Bonifacio and Preparata.

\section{Classical One-spin system.}
\label{classical}
\subsection{Stationary points and their stability}

Hence, we consider  the following Hamiltonian
\begin{equation}
H= 2 \epsilon  s^z  + \omega \bar{b} b + g \left( \bar{b} s^-+ b s^+ \right)
\label{bfconspinclas1}
\end{equation}
Here $s^z = s^3, s^\pm = s^1\pm i s^2$  are spins variables, and $b,\bar{b}$ is a  harmonic oscillator.
The Poisson brackets read
\begin{equation}
\{ s^a , s^b \} = - \epsilon_{abc} s^c, \quad \{ b , \bar{b} \} = i
\label{poisson1}
\end{equation}
Changing the sign of $g$ amounts to changing $(b,\bar{b})\to (-b,-\bar{b})$ which is a symplectic transformation. Rescaling the time we can assume $g=1$.

The Poisson bracket of the spin variables is degenerate. To obtain a symplectic manifold, we fix the value of the Casimir function
$$
\vec{s} \cdot \vec{s} = (s^z)^2 + s^+ s^- = s_{cl}^2
$$
so that the corresponding phase space becomes the product of a sphere by a plane and has a total dimension $4$.
Let us write
$$
H=  H_0 +  \omega H_1 
$$
with
\begin{equation}
H_1 =  \bar{b} b + s^z,\quad H_0 = 2\kappa s^z + \bar{b} s^-+ b s^+, \quad \kappa = \epsilon - \omega/2
\label{hamclas}
\end{equation}
Clearly we have 
$$
\{ H_0, H_1 \} = 0
$$
so that the system in integrable\footnote[1]{This remains true for the $N$-spin system. In the Integrable Community this model is known to be  a limiting case of the Gaudin model \cite{YKA}}. 
Hence we can solve simultaneously the evolution equations
$$
\partial_{t_0} f = \{ H_0, f \}, \quad \partial_{t_1} f = \{ H_1, f \}
$$
Once $f(t_0,t_1)$ is known, the solution of the equation of motion $\partial_{t} f = \{ H, f \}= (\partial_{t_0}+ \omega \partial_{t_1})f$ is given by $f(t,\omega t)$. The equations of motion read
\begin{align}
\partial_{t_1} b &= - ib     &\partial_{t_0} b &=  - i   s^-   &\partial_{t} b &=  - i   s^-  -i\omega b
   \label{motionb2} \\
\partial_{t_1} {s}^z &=0  & \partial_{t_0} {s}^z &= i  ( \bar{b} s^- - b s^+ ) & \partial_{t} {s}^z &=   i  ( \bar{b} s^- - b s^+ ) 
\label{motionsz2} \\
\partial_{t_1}{s}^+ &= i s^+  &\partial_{t_0}{s}^+ &= 2i \kappa s^+ -2i \bar{b} s^z &\partial_{t}{s}^+ &=   2i \epsilon s^+ -2i \bar{b} s^z  \label{motions+2}\\
\partial_{t_1}{s}^- &= -i  s^-  &\partial_{t_0}{s}^- &= -2i \kappa s^- +2i b s^z &\partial_{t}{s}^- &= -2i \epsilon s^- +2i b s^z  \label{motions-2}
\end{align}
The $t_1$ evolution is simply a simultaneous rotation around the $z$ axis of the spin and a rotation of the same angle
of the harmonic oscillator part: 
\begin{equation}
b(t_0,t_1) = e^{-it_1} b(t_0),\quad s^\pm(t_0,t_1) = e^{\pm it_1}s^\pm(t_0),\quad s^z(t_0,t_1) =  s^z(t_0)
\label{t1flow}
\end{equation}
In physical terms, the conservation of $H_1$ corresponds to the 
conservation of the total number of particles in the system, when this model (with $N$ quantum spins 1/2)
is used to describe the coherent dynamics between a Fermi gas and a condensate of 
molecules~\cite{Levitov}.

We wrote the full equations of motion to emphasize the fact that the points
\begin{equation}
s^\pm =  0, \quad s^z = \pm  s_{cl},\quad b=b^\dag = 0
\label{stationary1}
\end{equation}
are  {\em all} the critical points (or stationary points i.e. all time derivatives equal zero) of {\it both} time evolutions $t_0$ and $t_1$ and hence are very special. Any Hamiltonian, function of $H_0$ and $H_1$, will have these two points among its critical points. However, it may have more.

For instance when $H=H_0+\omega H_1$ an additional  family of stationnary points exist when $\omega^2 \epsilon^2 < s_{cl}^2$. They are  given by
\begin{equation}
s^{\pm} = \sqrt{s_{cl}^2 -\epsilon^2 \omega^2}\; e^{\pm i\varphi}, \quad
s^z = -\epsilon \omega, \quad b= -{1\over \omega} \sqrt{s_{cl}^2 -\epsilon^2 \omega^2}\; e^{-i\varphi}
\quad \forall \varphi
\label{stationary2}
\end{equation}
their energy is given by
$$
E' = -{1\over \omega}(s_{cl}^2 + \omega^2 \epsilon^2)
$$
The energies of the configurations Eq.~(\ref{stationary1}) are
$$
E = \pm 2 \epsilon s_{cl}
$$
so that 
$$
E-E' = {1\over \omega}(s_{cl} \pm \epsilon \omega)^2 >0
$$
and we see that $E'$ represents the degenerate ground states of $H$, breaking rotational invariance around the $z$ axis.

\bigskip

We are mostly interested in the configurations Eq.~(\ref{stationary1}).  To fix ideas we assume $\epsilon <0$ so that among these two  configurations, the one with minimal energy is  the spin up.  It looks like being a minimum however it becomes unstable for some values of the parameters.  To see it, we perform the analysis of the small fluctuations around this configuration.
We assume that $b, \bar{b}, s^{\pm}$ are first order and $s^z =  s_{cl} e + \delta s^z$ where $e=\pm 1$
according to whether the spin is up or down. Then 
$\delta s^z$ is determined by saying that the spin has length $s_{cl}$ 
$$
\delta s^z = - {e\over 2s_{cl}} s^- s^+
$$
This is of second order and is compatible with Eq.~(\ref{motionsz2}). The linearized equations of motion (with respect to $H$) are
\begin{eqnarray}
\dot{b} &=& -i \omega b - i  s^-  \label{motionb1}\\
\dot{s}^- &=& -2i \epsilon s^- +2is_{cl}  e b  \label{motions-1}
\end{eqnarray}
and their complex conjugate. 
We look for eigenmodes in the form
$$
b(t)=b(0)e^{-2iEt},\quad s^- = s^-(0) e^{-2iE t}
$$
 We get from Eq.~(\ref{motions-1})
$$
s^-(0) = -{s_{cl}} {e\over E-\epsilon} b(0)
$$
Inserting into Eq.~(\ref{motionb1}),
we obtain the self-consistency equation for $E$:
\begin{equation}
E = {\omega\over 2} - {s_{cl}\over 2}  {e \over E-\epsilon   } 
\label{eqE1}
\end{equation}
%or
%$$
%E^2 -\left(\epsilon + {\omega\over 2}\right) E + {1\over 2} (\omega \epsilon + s_{cl} e ) = 0
%$$
The discriminant of this second degree equation for $E$ is $ \kappa^2 - 2 s_{cl} e$.
If $e=-1$ it is positive and we have a local energy maximum. However  if $e=+1$, the roots become complex when $\kappa^2 \leq 2s_{cl}$.  In that case  one of the  modes exponentially  increases with time, i.e. the point is  {\em unstable}.
Since we are in a situation where $\kappa \leq 0$, the transition occurs when $\kappa = - \sqrt{2s_{cl}}$.

\begin{figure}[hbtp]
\begin{center}
\includegraphics[height= 10cm]{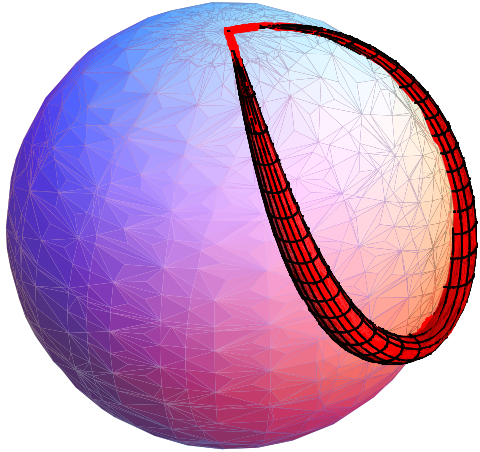}  
\caption{The image in phase space of the level set of the critical unstable point $H_0=2\kappa s_{cl}, H_1=s_{cl}$ is a pinched two dimensional torus.}
\label{figphase3D}
\end{center}
\nonumber
\end{figure}

The level set of the unstable point, i.e. the set of points in the four-dimensional phase space of the spin-oscillator system,  which have the same values of $H_0$ and $H_1$ as the critical point  $H_0=2\kappa s_{cl}, H_1=s_{cl}$ has the topology of a pinched two-dimensional torus see Fig.[\ref{figphase3D}]. This type of stationary point is known in the mathematical litterature as a 
focus-focus singularity~\cite{Zou92,Zung97,Cushman01}. 
The above perturbation analysis shows that in the immediate vicinity of the critical point, the pinched
torus has the shape of two cones that meet precisely at the critical point. One of these cones is associated to
the unstable small perturbations, namely those which are exponentially amplified, whereas the other cone
corresponds to perturbations which are exponentially attenuated. Of course, these two cones are connected, so that
any initial condition located on the unstable cone gives rise to a trajectory which reaches eventually 
the stable cone in a finite time. Note that this longitudinal motion from one cone to the other is 
superimposed to a spiraling motion around the closed cycle of the pinched torus which is generated by the action
of $H_1$.

\begin{figure}[hbtp]
\begin{center}
\includegraphics[height= 7cm]{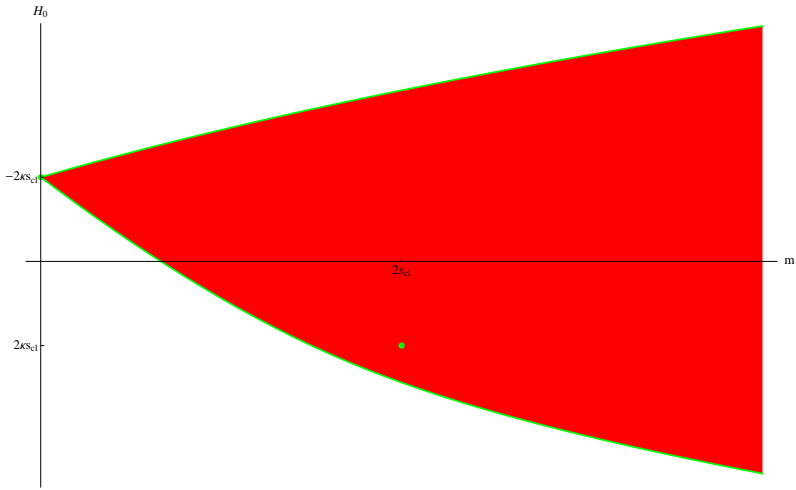}  
\caption{The image in $R^2$ of the phase space by the map $(H_1\equiv m-s_{cl},H_0)$. The green dots are the critical points. The stable one is the point on the vertical axis $(m=0,-2\kappa s_{cl})$ and the unstable one is located at $(m=2s_{cl}, 2\kappa s_{cl})$. The upper and lower green boundaries are obtained as explained in the text. }
\label{figmoment}
\end{center}
\nonumber
\end{figure}

An important object is the image of phase space into $R^2$ under the map $(H_1,H_0)$. It is shown in Fig.[\ref{figmoment}]. It is a convex domain in $R^2$. The upper and lower green boundaries are obtained as follows. We set 
$ x= \bar{b} b$ and 
\begin{equation}
H_1 = x + s^z \equiv m-s_{cl}, \quad {\rm Max}(0,m-2s_{cl}) \leq x \leq m
\label{xvariation}
\end{equation}
 where the parameter $m$ is introduced to ease the comparison with the quantum case. Then we set
 $$
 b = \sqrt{x}e^{i\theta}, \bar{b} = \sqrt{x}e^{-i\theta},\quad s^{\pm} = \sqrt{s_{cl}^2 - (s^z)^2} e^{\mp i\varphi}
= \sqrt{(m-x)(2s_{cl}-m+x)}e^{\mp i\varphi}
 $$
 so that 
 $$
 H_0= 2\kappa(m-s_{cl}-x) + 2 \sqrt{x(m-x)(2s_{cl}-m+x)} \cos(\theta-\varphi)
 $$
It follows that
$$
 2\kappa(m-s_{cl}-x) - 2 \sqrt{x(m-x)(2s_{cl}-m+x)} \leq H_0 \leq 2\kappa(m-s_{cl}-x) + 2 \sqrt{x(m-x)(2s_{cl}-m+x)}
 $$
 To find the green curves in Fig.[\ref{figmoment}] we have to minimize the left hand side and maximize the right hand side in the above inequalities when $x$ varies in the bounds given in Eq.~(\ref{xvariation}).

\bigskip

In the unstable case, because an initial condition close to the critical point leads to a large
trajectory along the pinched torus, it is interesting to know 
the complete solutions of the equations of motion. They are easily obtained as follows. Remark that 
\begin{eqnarray*}
(\dot{s}^z )^2 &=& - ( \bar{b} s^- - b s^+ )^2 = - ( \bar{b} s^- + b s^+ )^2
+4  \bar{b} b s^+ s^- \\
&=& - (H_0-2\kappa s^z)^2 +4  (H_1-s^z )(s_{cl}^2-(s^z)^2)
\end{eqnarray*}
or
$$
(\dot{s}^z )^2 = 4 (s^z)^3 -4(H_1 +\kappa^2) (s^z)^2 + 4(\kappa H_0 - s_{cl}^2) s^z + (4s_{cl}^2 H_1 -H_0^2)
$$
Hence, $s^z$ satisfies an equation of the Weierstrass type and is solved by elliptic functions.
Indeed, setting
$$ 
s^z = {1\over 3}( H_1+\kappa^2) -  X, \quad \dot{s}^z = i Y
$$
the equation becomes
\begin{equation}
Y^2 = 4 X^3 - g_2 X - g_3
\label{ellipticcurve}
\end{equation}
with
$$
g_2 = {4\over 3}  ( H_1^2 - 3 H_0 \kappa  + 2  H_1 \kappa ^2 +  \kappa ^4 + 3  s_{cl}^2)
$$
$$
g_3 = {1 \over 27} (-27  H_0^2 - 8  H_1^3 + 36  H_0 H_1 \kappa  - 24  H_1^2 \kappa ^2 + 
   36  H_0 \kappa ^3 - 24  H_1 \kappa ^4 - 8  \kappa ^6 + 72  H_1 s_{cl}^2 - 
   36  \kappa ^2 s_{cl}^2)
 $$
Therefore, the general solution of the one-spin system reads
\begin{equation}
s^z = {1\over 3}( H_1+\kappa^2) -  \wp(it + \alpha)
\label{fullsolutionsz}
\end{equation}
and
\begin{equation}
\bar{b} b = H_1 - s^z =  {1\over 3} (2 H_1 - \kappa^2) + \wp(it+ \alpha) 
\label{fullsolutionb2}
\end{equation}
where $\alpha $ is an integration constant and $\wp(\theta)$ is the Weierstrass function associated to the curve Eq.~(\ref{ellipticcurve}).
Initial conditions can be  chosen such that when $t=0$, we start from a point intersecting the real axis
$Y=0$. This happens when $\alpha $ is half a  period:
$$
 \alpha = \omega_1~{\rm or}~ \omega_2~{\rm or}~  \omega_3 = - \omega_1 - \omega_2
$$

\bigskip

This general solution however is not very useful because the physics we are interested in
lies on Liouville tori  specified by very particular values of the conserved quantities $H_0, H_1$.
For the configuration Eq.~(\ref{stationary1}) with spin up,  the values of the Hamiltonians are
$$
H_0 = 2\kappa s_{cl}, \quad H_1 = s_{cl}
$$
In that case we have
$$
g_2 = {4 \over 3} \Omega^4,\quad g_3 = {8 \over 27} \Omega^6
$$
and
$$
4 X^3 - g_2 X - g_3 = 4\left(X + {1\over 3}\Omega^4 \right)^2\left(X - {2\over 3}\Omega^2 \right)
$$
where we have defined
\begin{equation}
\Omega^2 = 2 s_{cl} - \kappa^2 >0 {\rm ~~in~the~} unstable~{\rm case.}
\label{defomega}
\end{equation}
Hence, we are precisely in the case where the elliptic curve degenerates. Then the solution of the equations of motion is expressed in terms of trignometric functions
$$
X(t) = {2\over 3} \Omega^2 -\Omega^2\tanh^2 (\Omega (t-t_0) )
$$
and
\begin{equation}
x(t) \equiv \bar{b} b(t)  = {\Omega^2
\over \cosh^2\Omega(t-t_0)}
\label{bbarclas}
\end{equation}

This solution represents a single solitonic pulse centered at time $t_0$. When the initial
condition is close but not identical to 
the unstable configuration with spin up and $b=\bar{b}=0$, this unique soliton is replaced
by a periodic sequence of pulses that can be described in terms of the Weiertrass
function as shown above. Note that in the context of cold fermionic atoms,
$\bar{b}b$ represents the number of molecular bound-states that have been formed in the system.

\subsection{Normal form and monodromy}
\label{secclassicalmonodromy}

The dynamics in the vicinity of the unstable equilibrium can also be visualized
by an appropriate choice of canonical variables in which the quadratic parts in the
expansions of $H_0$ and $H_1$ take a simple form. Let us introduce the angle $\nu \in ]0,\pi/2[$ 
such that $|\kappa|=\sqrt{2s_{cl}}\cos\nu$ and $\Omega=\sqrt{2s_{cl}}\sin\nu$. As we
have shown, instability occurs when $\kappa^{2}<2s_{cl}$, which garanties that $\nu$ is real.
Let us then define two complex coordinates $A_s$ and $A_u$ by:
\begin{equation}
\left(\begin{array}{c} A_{s} \\ A_{u} 
\end{array}\right)= \frac{1}{\sqrt{2\sin\nu}}
\left(\begin{array}{cc} e^{-i\nu/2} & e^{i\nu/2} \\
e^{i\nu/2} & e^{-i\nu/2} 
\end{array}\right) 
\left(\begin{array}{c} b \\ \frac{s^{-}}{\sqrt{2s_{cl}}} 
\end{array} \right)
\end{equation}
The classical Poisson brackets for these variables are: 
\begin{eqnarray}
\{A_s,\bar{A}_{s}\} & = & \{A_u,\bar{A}_{u}\} = 0 \\
\{A_s,\bar{A}_{u}\} & = & \{\bar{A}_{s},A_{u}\} = 1
\end{eqnarray} 
Here we have approximated the exact relation $\{s^{+},s^{-}\}=2is^{z}$ by $\{s^{+},s^{-}\}=2is_{cl}$,
which is appropriate to capture the linearized flow near the unstable fixed point.
After subtracting their values at the critical point which do not influence the dynamics, 
the corresponding quadratic Hamiltonian $H_0$ and global rotation generator $H_1$ read:
\begin{eqnarray}
H_0 & = & 2\Omega \Re(\bar{A}_{s}A_{u})+ \kappa H_1 \\
H_1 & = & 2\Im(\bar{A}_{s}A_{u}) 
\end{eqnarray}
With the above Poisson brackets, the linearized equations of motion take the form:
\begin{eqnarray}
\dot{A}_{s} & = & (-\Omega - i \kappa ) A_{s} \label{evolAs}\\
\dot{A}_{u} & = & (\Omega  - i \kappa) A_{u} \label{evolAu}
\end{eqnarray}
With these new variables, the pinched torus appears, in the neighborhoood
of the unstable point, as the union of two planes intersecting transversally. 
These are defined by $A_{u}=0$ which corresponds to the stable branch, and
by $A_{s}=0$ which gives the unstable branch. The global rotations generated
by $H_1$ multiply both $A_{s}$ and $A_{u}$ by the same phase factor $e^{-it_1}$. We may then
visualize these two planes as two cones whose common tip is the critical point, as
depicted on Fig.~\ref{figphase3D}. The stable
cone is obtained from the half line where $A_{s}$ is real and positive and $A_{u}=0$,
after the action of all possible global rotations. A similar description holds for the
unstable cone. As shown by Eq.~(\ref{bbarclas}), 
any trajectory starting form the unstable branch eventually reaches the stable 
one. This important property will play a crucial role below in the computation
of the monodromy.

For latter applications, specially in the quantum case, it is useful to write down
explicitely $A_{s}$ and $A_{u}$ in terms of their real and imaginary parts:
\begin{eqnarray}
A_{s} & = & (P_{1}-iP_{2})/\sqrt{2} \\
A_{u} & = & (X_{1}-iX_{2})/\sqrt{2}
\end{eqnarray}
This reproduces the above Poisson brackets, provided we set $\{X_{i},X_{j}\}=0$, 
$\{P_{i},P_{j}\}=0$, $\{P_{i},X_{j}\}=\delta_{ij}$ for $i,j\in\{1,2\}$.
The quadratic Hamitonian now reads:
\begin{eqnarray}
H_0 & = & \Omega (X_{1}P_{1}+X_{2}P_{2}) + \kappa H_1 \\
H_1 & = & X_{1}P_{2}-X_{2}P_{1}
\end{eqnarray}
which is the standard normal form for the focus-focus 
singularity~\cite{Zou92,Zung97,Cushman01}.

We are now in a position to define and compute the monodromy attached to a closed path
around the unstable point in the $(H_0,H_1)$ plane as in Fig.~\ref{figmoment}.
Let us consider a one parameter family of regular invariant tori
which are close to the pinched torus. This family can be described by
a curve in the $(H_0,H_{1})$ plane, or equivalently, in the complex plane
of the $\bar{A}_{s}A_{u}$ variable. Let us specialize to the case of a closed loop.
Its winding number around the critical value in the $(H_0,H_{1})$ plane is the same 
as the winding number of  $\bar{A}_{s}A_{u}$ around the origin in the complex plane.
Let us consider now the path $\chi\in[0,2\pi]\rightarrow\bar{A}_{s}A_{u}=\eta^{2}e^{i\chi}$
where $\eta>0$ is assumed to be small. To construct the monodromy, we need to define,
for each value of $\chi$, two vector fields on the corresponding torus, which are
generated by $\chi$-dependent linear combinations of $H_0$ and $H_1$ and whose flows
are $2\pi$-periodic. Furthermore we require that the periodic orbits 
of these two flows generate a cycle basis $(e_1,e_2)$ in the two-dimensional homology of the torus.
This basis evolves continuously as $\chi$ increases from 0 to $2\pi$. The monodromy
expresses the fact that $(e_1,e_2)$ can turn into a different basis after one closed loop
in the $(H_0,H_{1})$ plane. The discussion to follow has been to a large extent inspired by
Cushman and Duistermat~\cite{Cushman01}. 

One of these two flows can be chosen as the $2\pi$ global rotation generated by $H_1$.
The corresponding orbits are circles which provide the first basic cycle $e_1$.
For the other one, we first emphasize that the flow generated by
$H_0$ is not periodic in general. So, to get the other basic cycle $e_2$, we have to consider
the flow generated by an appropriate linear combination of $H_0$ and $H_1$. 
To construct it, it is convenient to consider an initial
condition $(A_{s}(0),A_{u}(0))$ such that $\bar{A}_{s}(0)A_{u}(0)=\eta^{2}e^{i\chi}$ and 
$|A_{s}(0)| \ll |A_{u}(0)|$, that is close to the unstable manifold of the pinched torus.
Let us pick a small enough $\zeta>0$ so that the linearized equations of motion are
still acurate when $|A_{s}(0)|^{2}+|A_{u}(0)|^{2}<\zeta^{2}$, and let us also assume that
$\eta \ll \zeta$. Using the global rotation invariance of the dynamics, we may choose
$A_{u}(0)=\zeta$ and $A_{s}(0)=\frac{\eta^{2}}{\zeta}e^{-i\chi}$. If we let the system evolve
starting from this initial condition, $|A_{s}|$ will first decrease and $|A_{u}|$  
will increase, so the trajectory becomes closer to the unstable torus, along its
unstable manifold. After a finite time $t_{1}$, the trajectory reappears in the neighborhood
of the unstable fixed point, but now, near the stable manifold. This behavior can be seen
either as a consequence of Eq.~(\ref{bbarclas}) for the trajectories on the pinched torus
and extended to nearby trajectories by a continuity argument, or can be checked directly from
the explicit solution of the classical dynamics Eqs.~(\ref{fullsolutionsz},\ref{fullsolutionb2})
in terms of the Weierstrass function. We may thus choose $t_1$ 
such that $|A_{s}(t_1)|=\zeta$. A very important property of $A_{s}(t_1)$ is that it has
a well defined limit when $\eta$ goes to zero, because in this limit, one recovers the 
trajectory on the pinched torus such that $A_{s}(0)=0$ and $A_{u}(0)=\zeta$. 
As a result, when $\eta$ is small enough, 
the argument of $A_{s}(t_1)$ in polar coordinates weakly depends on $\chi$ and
the winding number of $A_{s}(t_1)$ when $\chi$ goes from 0 to $2\pi$ vanishes. 
We then let the system evolves until time $t_2$ using the linearized flow.
The time $t_2$ is chosen in order to recover $|A_{u}(t_2)|=\zeta$. Using Eq.~(\ref{evolAu})
and the fact that $\bar{A}_{s}A_{u}$ is conserved,
this gives $t_{2}-t_{1}=(2/\Omega)\ln(\zeta/\eta)$. Furthermore, from  
$\bar{A}_{s}(t_1)A_{u}(t_1)=\eta^{2}e^{i\chi}$ and  Eq.~(\ref{evolAu}), we deduce:
\begin{equation}
A_{u}(t_2)=e^{i\chi}\exp\left(-2i{\kappa \over \Omega}\ln\frac{\zeta}{\eta}\right)A_{s}(t_1)
\end{equation}
In general, we have no reason to expect that $A_{u}(t_2)$ should be equal to $A_{u}(0)$ so,
to get a periodic flow on the torus, the evolution generated by $H_0$ during the time $t_2$
has to be followed by a global rotation of angle $\beta (\chi)$ given by:
\begin{equation}
e^{i\beta (\chi)}=e^{i\chi}\exp\left(-2i{\kappa \over \Omega} \ln\frac{\zeta}{\eta}\right)
A_{s}(t_1)/\zeta
\end{equation}
The sign of $\beta$ reflects the fact that the $H_1$ flow applied during the time $\beta$
multiplies $A_{s}$ and $A_{u}$ by $e^{-i\beta}$.
Note that because the flows associated to $H_0$ and $H_1$ commute, the composition
of the $H_0$-flow during time $t_2$ and of the $H_1$ flow along an angle $\beta$ can also
be viewed as the flow generated by the linear combination $H_0+(\beta/t_2)H_1$ during time $t_2$.
We have established the periodicity of this flow for the orbit starting from
$A_{u}(0)=\zeta$ and $A_{s}(0)=\frac{\eta^{2}}{\zeta}e^{-i\chi}$. But any other orbit on the
invariant torus charaterized by $\bar{A}_{s}A_{u}=\eta^{2}e^{i\chi}$ can be deduced from this
one by a global rotation, which commutes with the flow of $H_0+(\beta/t_2)H_1$. This establishes
the periodicity of this latter flow, and therefore allows us to construct the other basic cycle
$e_2(\chi)$, which depends smoothly on $\chi$.

Because the winding number of $A_{s}(t_1)$ vanishes provided $\eta$ is small enough, we have the
crucial relation:
\begin{equation}
\beta (2\pi)=\beta (0)+2\pi
\end{equation}
In other words, when we follow by continuity the periodic flow generated by 
$H_0+(\beta/t_2)H_1$ during time $t_2$, the final flow is deduced from the initial one
by a global rotation of $2\pi$. 
To summarize, as $\chi$ increases smoothly from $0$ to $2\pi$, $e_1$ is left unchanged, whereas $e_2$
evolves into $e_2+e_1$. In this basis, the monodromy matrix of the focus-focus singularity
is then:
\begin{equation}
M=\left(\begin{array}{lr} 1 & 1 \\ 0 & 1 \end{array} \right)
\end{equation} 

\subsection{Reduced system.}

\begin{figure}[hbtp]
\begin{center}
\includegraphics[height= 5cm]{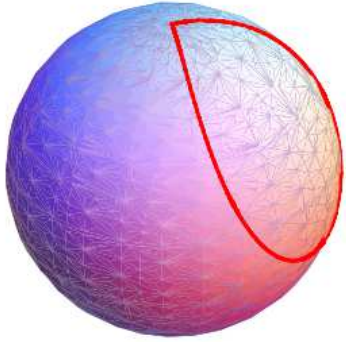}  
\caption{The sphere represents the reduced phase-space obtained by fixing the conserved quantity $H_1$ to its value $s_{cl}$
at the critical point. Here, this point corresponds to the north pole on the sphere. The red curve is the image of the pinched torus
after the reduction procedure. It is obtained by setting the other integral of motion $H_0$ to its critical value $H_0=2\kappa s_{cl}$.
For this, we have used the expression of $H_0$ given in Eq.~(\ref{H0reducedclassical}). The singular nature of the pinched
torus is reflected by the cusp of the red curve at the north pole.}
\label{phaseportrait3D1}
\end{center}
\nonumber
\end{figure}

As we  have discussed before, the common level set of $H_0$ and $H_1$ which contains the unstable critical point
is a pinched torus. Since we are mostly interested in the time evolution of the oscillator energy $\bar{b}b$, and
because this quantity is invariant under the Hamiltonian action of $H_1$, it is natural to reduce the dynamics
to the orbits of $H_1$. For an initial condition on the pinched torus, this amounts to discard the spiraling 
motion around the torus, and to concentrate on the longitudinal motion from the stable cone to the
unstable one. In other words, this reduction procedure reduces the pinched torus into a curve which has a cusp
at the critical point. This is illustrated by the thick curve on Fig.~\ref{phaseportrait3D1}.

\begin{figure}[hbtp]
\begin{center}
\includegraphics[height= 6cm]{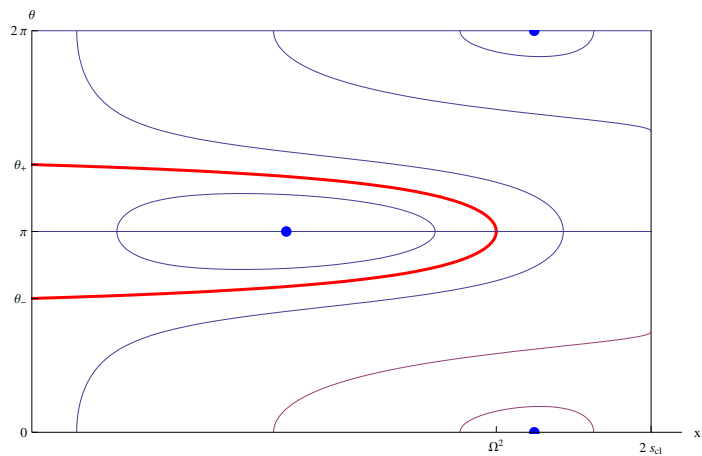}  
\caption{The phase portrait corresponding to the Hamiltonian $H_0$, Eq.~(\ref{H0reducedclassical}), in the $(\theta,x)$ variables, which
are natural coordinates for the reduced system. 
The thick red line corresponds to the critical separatrix $H_0= 2\kappa s_{cl}$, which is the image of the
pinched critical torus after symplectic reduction. Note that the unstable critical point is mapped onto the
vertical axis at $x=0$, so that the critical separatrix is indeed a loop.   
The blue dots correspond to  the minimal and maximal values of $H_0$ when $H_1= s_{cl}$.}
\label{figphase}
\end{center}
\nonumber
\end{figure}

An other description of this trajectory is given by the thick curve on Fig.~\ref{figphase} that will be discussed below.
In practice, the first thing to do to implement this reduction is to fix the value of $H_1$ to its critical value
\begin{equation}
H_1 = \bar{b} b + s^z = s_{cl}
\label{H1critical}
\end{equation}
This defines a submanifold ${\cal H}_{s_{cl}}$ in phase space. Since $H_0$ and $H_1$ Poisson commute we may consider the reduced system  
obtained by performing a Hamiltonian reduction by the one parameter group  generated by $H_1$. 
The reduced phase space is 
the  quotient of ${\cal H}_{s_{cl}}$ by the flow generated by $H_1$ (the trivial $t_1$  phases in Eq.~(\ref{t1flow})). It is of dimension 2.
Explicitly, we set: 
$$
x =  \bar{b}b >0 \Longrightarrow  b = \sqrt{x} e^{i\theta},\quad \bar{b} =   \sqrt{x} e^{-i\theta}
$$
From Eq.~(\ref{H1critical}) and  from the condition $(s^z)^2+ s^+  s^- = s_{cl}^2$ we deduce
\begin{equation}
s^z = s_{cl}-x, \quad {\rm and} \quad  s^\pm = \sqrt{ x(2s_{cl}-x)} e^{\mp i\varphi}
\label{sreduit}
\end{equation}
the reduced Hamiltonian $H_0$ reads
\begin{equation}
H_0 = 2\kappa s_{cl} + 2x \sqrt{2s_{cl} -x} \cos(\theta-\varphi) - 2\kappa x
\label{H0reducedclassical}
\end{equation}
Notice that $x$ and $\theta-\varphi$ are invariant by the $H_1$ flow, and they can be taken as coordinates on the reduced phase space. These coordinates are canonically conjugate as we easily see by writing the symplectic form:
$$
\omega = -i \delta b \wedge \delta \bar{b} + i {\delta s^+ \over s^+} \wedge \delta s^- =
\delta x \wedge \delta  (\theta-\varphi)
$$
In the following, when talking about this reduced system, we will simply set $\varphi=0$. These coordinates are very convenient but one has to be aware that they are singular at $x=0$ and $x=2s_{cl}$. The whole segment  $x=0, 0 \leq \theta < 2\pi$ should be identified to one point and similarly for the segment $x=2s_{cl}, \, 0 \leq \theta < 2\pi$.

The equations of motion of the reduced system read
\begin{eqnarray*}
\dot{x} &=& 2 x \sqrt{2s_{cl} -x} \sin \theta \\
\dot{\theta} &=& -2\kappa +2 \left( \sqrt{2s_{cl} -x}  - {x\over 2  \sqrt{2s_{cl} -x}} \right)\cos \theta
\end{eqnarray*}
From this we see that the critical points are given by
$$
x=0,\quad  \cos \theta = {\kappa \over \sqrt{2s_{cl}}},\quad {\rm or}\quad z^2 - {2\kappa  \over \sqrt{2s_{cl}}} z + 1 =  0, \quad z=e^{i\theta}
$$
In our case $\kappa$ is negative, so that the solutions of the quadratic equations will be defined as
$$
z_\pm  = {\kappa \pm i \Omega \over \sqrt{2s_{cl}}} = - e^{\mp i\nu} , \quad \Omega^2 =  2s_{cl} - \kappa^2, \quad 0 \leq \nu \leq \pi/2
$$
These points exist precisely in the unstable regime $\kappa^2 \leq 2s_{cl}$. They correspond to only one point in phase space  whose energy is given by
\begin{equation}
H_0= 2\kappa s_{cl} \equiv E_{c}
\label{defEc}
\end{equation}
hence they correspond to the unstable point. The image of the pinched critical torus after the reduction procedure
is shown on Fig.~\ref{figphase} as the thick line connecting $z_{+}$ and $z_{-}$. Note that this line intersects
the $x=0$ segment with a finite angle. This angle reflects precisely the pinching of the torus at the 
unstable critical point in the original four-dimensional phase space.

\bigskip

Another critical point is obtained by setting $\theta = \pi$. Then 
\begin{equation}
x = 2s_{cl} - X^2, \quad {\rm with } \quad X={1\over 3} (-\kappa + \sqrt{\kappa^2 + 6 s_{cl}} )
\label{groundstate}
\end{equation}
Since $0\leq x \leq 2s_{cl}$, we should also have $0 \leq X^2 \leq 2s_{cl}$ which is the case 
when $\kappa \geq -\sqrt{2s_{cl}} $, that is to say in the unstable region. 
The energy is given by
\begin{equation}
H_0= 2\kappa s_{cl} -2(X+\kappa) (2s_{cl} -X^2)
\label{ground0clas}
\end{equation}
We see from that equation that, in the unstable region, $H_0-2\kappa s_{cl}$  is always negative and the configuration Eq.~(\ref{groundstate}) represents the ground state.

\bigskip

Finally, the energy maximum of this reduced system is obtained for $\theta=0$ and $x=2s_{cl}-Y^2$,
with $Y=(\kappa+\sqrt{\kappa^{2}+6})/3$. On the phase portrait shown on Fig.~\ref{figphase}, one sees
a curve which contains the point at $x=2s_{cl}$ and which looks like a separatrix. However, unlike the situation
around $x=0$, we notice that it does not correspond to any stationary point of the unreduced Hamiltonian. In fact it is simple to show that the curve is actually  tangent to the vertical line $x=2s_{cl}$. On the sphere, this is the trajectory passing through the south pole.

\section{Quantum One-spin system.}
\label{quantique}
\subsection{Energy spectrum}

We now consider the quantum system. We set as in the classical case
$$
H = H_0+\omega H_1
$$
with
\begin{eqnarray*}
H_0 &=& (2\epsilon -\omega) s^z +   b^\dag s^- + b s^+
= 2\kappa s^z + b^\dag s^- + b s^+, \quad\quad H_1=  b^\dag b + s^z 
\end{eqnarray*}
where  as before
\begin{equation}
\kappa = \epsilon - \omega/2
\label{defkappa}
\end{equation}
We impose the commutation relations
$$
[b,b^\dag ] = \hbar, \quad [s^+,s^-] = 2\hbar s^z,\quad [s^z,s^\pm] = \pm \hbar  s^\pm
$$
We assume that the spin acts on a spin-$s$ representation.
$$
s^z \vert m \rangle = \hbar m \vert m \rangle,\quad
s^\pm \vert m \rangle = \hbar  \sqrt{s(s+1) - m(m \pm 1)}  \; \vert m \pm 1\rangle,
\quad m = -s, -s+1, \cdots , s-1,s
$$
where $2s$ is integer.  Of course
$$
(s^z)^2 +{1\over 2} (s^+ s^- + s^- s^+) = \hbar^2 s(s+1)
$$
We still have
$$
[ H_0, H_1] = 0
$$
and the quantum system is integrable. The Hamiltonian $H$ can be diagonalized by Bethe Ansatz, but we will not follow this path here.  For related studies along this line for the  Neumann model see  \cite{BeTa}. 
Note that recently, the non equilibrium dynamics of a similar quantum integrable model (the Richardson model) has been studied by a combination
of analytical and numerical tools.~\cite{Faribault09}  
Let:
\begin{equation}
e_n = (2b^\dag)^{n} \vert 0 \rangle \otimes (s^+)^{M-n} \vert -s \rangle, \quad {\rm Sup}(0,M-2s)  \leq n \leq  M
\label{defen}
\end{equation}
For all these states one has:
$$
H_1 e_n = \hbar (M-s) e_n
$$
Since $H_0$ commutes with $H_1$ we can restrict $H_0$ to the subspace spanned by the $e_n$.
Hence, we write:
$$
\Psi= \sum_{n= {\rm Sup}(0,M-2s) }^{M}  p_n \;\; {e_n\over ||e_n||}, \quad
$$
where the norm  $||e_n|| $ of the vector $e_n$ is given by:
$$
 ||e_n||^2 = \hbar^{2M-n}2^{2n}  {(2s)! (M-n)! n! \over (2s-M+n)!}
$$
Using
$$
H_0 e_n = 2\hbar \kappa (M-s-n) e_n +{ \hbar^2 \over 2} (M-n)(2 s +1-M+ n) e_{n+1} + 2\hbar n e_{n-1}
$$
the Schr\"odinger equation 
$$
i\hbar {\partial \Psi \over \partial t} = H_0 \Psi
$$
becomes:
\begin{eqnarray*}
i\hbar {\partial p_n \over \partial t}&=& \hbar^{3\over 2} \sqrt{(n+1)(2s+1-M+n)(M-n)} p_{n+1}  + \hbar^{3\over 2} \sqrt{n(2s-M+n)(M+1-n)} p_{n-1} \\
&& + 2\hbar \kappa (M-n-s)  p_n 
\end{eqnarray*}
and the eigenvector  equation $H_0 \Psi= E \Psi$ reads:
\begin{eqnarray}
 \hbar^{3\over 2} \sqrt{(n+1)(2s+1-M+n)(M-n)} p_{n+1}  &+& \hbar^{3\over 2} \sqrt{n(2s-M+n)(M+1-n)} p_{n-1} \\ 
 &+& 2\hbar \kappa (M-n-s)  p_n  =  E p_n
\label{EigEqM}
\end{eqnarray}
In this basis, the Hamiltonian is represented by a symmetric Jacobi matrix and  it  is easy to diagonalize it numerically. Varying $M=0,1,\cdots$
we construct the lattice of the joint spectum of $H_0,H_1$.
\begin{figure}[hbtp]
\begin{center}
\includegraphics[height= 7cm]{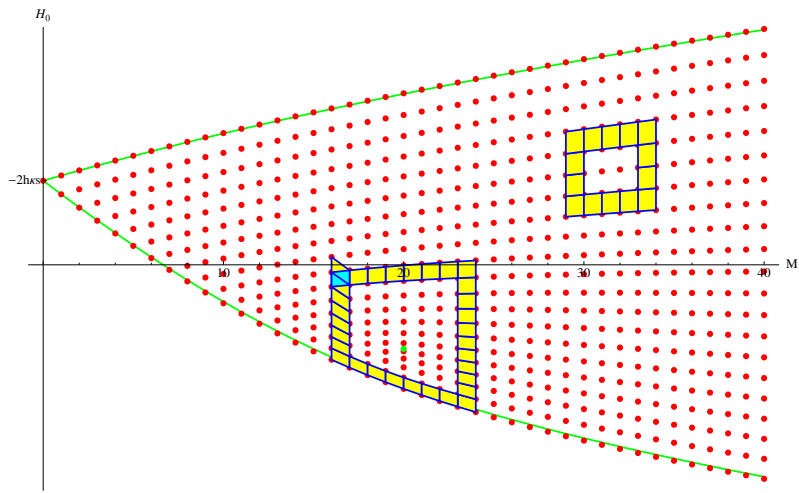}  
\caption{The lattice of the joint spectrum of $H_0$ and $H_1$ (or equivalently the integer $M$) in the unstable regime. 
The green point represents the classical unstable point ($H_1= s_{cl}$, or $M=2s$, and $H_0=2\kappa s_{cl}$). 
The parameters are $\hbar^{-1} = s=10$,  and $\kappa= -0.5\sqrt{2s_{cl}}$. 
This figure demonstrates the non-trivial quantum monodromy in this system. For this purpose, we follow by continuity 
an elementary cell of this lattice along two closed paths. When the path encloses the image of the critical
point in the  ($H_0$, $H_1$) plane, we see that an initial square cell does not preserve its shape at the
end of the cycle, but it evolves into a parallelogram with a tilt. In other words, the lattice of joint eigenvalues
exhibits a dislocation located at the critical value of $H_0$ and $H_1$.}
\label{figlattice}
\end{center}
\nonumber
\end{figure}
We see on Fig.~\ref{figlattice} that the lattice has a defect located near the unstable classical point. 
This defect induces a quantum monodromy when we transport a cell of the lattice around it. 
The rule to transport a cell is as follows. We start with a cell and we choose the next cell 
in one of the four possible positions (east, west, north, south) in such a way that two edges 
of the new cell prolongate two sides of the original cell, and the two cells have a common edge. 
We apply these rules on a path which closes. If the path encloses the unstable point, 
the last cell is different from the initial cell. 

The precise form of this lattice defect can be related to the classical monodromy matrix $M$,
as shown by V\~u Ng\d{o}c~\cite{San99}. Choosing locally a cycle basis $e_i$ on the Arnold-Liouville
tori is equivalent to specifying local action-angle coordinates $(J_i,\phi_i)$ such that $e_i$ is
obtained from the periodic orbit generated by $J_i$. After one closed circuit in the base space
of this fibration by tori, the basic cycles are changed into $e'_{i}=\sum_{j}M_{ji}e_{j}$ and therefore,
the new local action variables $J'_{i}$ are deduced from the initial ones by:
\begin{equation}
J'_{i}=\sum_{j=1}^{n}M_{ji}J_{j}
\label{changeactionvar}
\end{equation}
Heuristically, the Bohr-Sommerfeld quantization principle may be viewed as the requirement that
the quantum wave functions should be $2\pi$ periodic in each of the $n$ phase variables $\phi_i$,
which implies that $J_i$ should be an integer multiple of $\hbar$. When $\hbar$ is small, the discrete 
set of common eigenvalues of the conserved operators has locally the appearance of a regular lattice,
whose basis vectors $v_1$,...,$v_n$ are approximately given by: $dJ_{i}(v_{j})=\hbar \delta_{ij}$.
Because of the monodromy, this lattice undergoes a smooth deformation as one moves around a critical
value of the mutually commuting conserved quantities. After one complete turn, the basic lattice vectors  
are changed into  $v'_1$,...,$v'_n$, which are also approximately given by: $dJ'_{i}(v'_{j})=\hbar \delta_{ij}$.
Taking into account the definition of the classical monodromy matrix in Eq~(\ref{changeactionvar}), this
leads to the semi-classical monodromy on the joint spectrum:
\begin{equation}
v'_{i}=\sum_{j=1}^{n}(M^{-1})_{ij}v_{j}
\end{equation}
The discussion above was performed in the action angle variables $(J_1,J_2)$.  In the $(H_0,H_1)$ variables 
the basis $v, v'$ should be transformed by the Jacobian matrix of the mapping $(J_1,J_2)\to (H_0(J_1,J_2),H_1(J_1,J_2))$.

On Fig.~\ref{figlattice}, we see that after a clockwise turn around the singular value in the
$(H_1,H_0)$ plane (which corresponds to a positive winding in the $(H_0,H_1)$ plane), the lattice
vectors $v_M$ and $v_{H_{0}}$ are transformed into $v_M-v_{H_{0}}$ and $v_{H_{0}}$ respectively. This is
in perfect agreement with the classical monodromy matrix computed in section~\ref{secclassicalmonodromy}.

\bigskip

When $M=2s$, i.e. when $H_1$ takes its critical value corresponding to the unstable point,
$$
H_1 e_n  = \hbar s \; e_n
 $$
the Schr\"odinger equation simplifies to:
\begin{equation}
i\hbar {\partial p_n \over \partial t}= \hbar^{3\over 2} (n+1)\sqrt{(2s-n)} p_{n+1}  + \hbar^{3\over 2}n \sqrt{(2s+1-n)} p_{n-1} 
 + 2\hbar \kappa (s-n)  p_n 
 \label{schroerduite}
\end{equation}
Setting 
$$
x=\hbar n,\quad s_{cl}=\hbar s
$$
this equation becomes:
\begin{equation}
i\hbar {\partial p(x) \over \partial t}= (x+\hbar)\sqrt{(2s_{cl}-x)} p(x+\hbar) +x \sqrt{(2s_{cl}+\hbar-x)} p(x-\hbar)  + 2 \kappa (s_{cl}-x)  p(x)
\label{schroex}
\end{equation}
Introducing the shift operator
$$
e^{i\theta} p(x) =  p(x+\hbar)
$$
this can be rewritten as:
\begin{equation}
i\hbar {\partial p(x) \over \partial t}= \Big[\sqrt{(2s_{cl}-x)} e^{i\theta} x +x  
e^{-i\theta} \sqrt{(2s_{cl} -x)} + 2 \kappa (s_{cl}-x) \Big] p(x)
\label{defH0quant}
\end{equation}
and we recognize  the quantum version of the classical reduced Hamiltonian Eq.~(\ref{H0reducedclassical}), but with a very 
specific ordering of the operators.

\bigskip

\begin{figure}[hbtp]
\begin{center}
\includegraphics[height= 3cm]{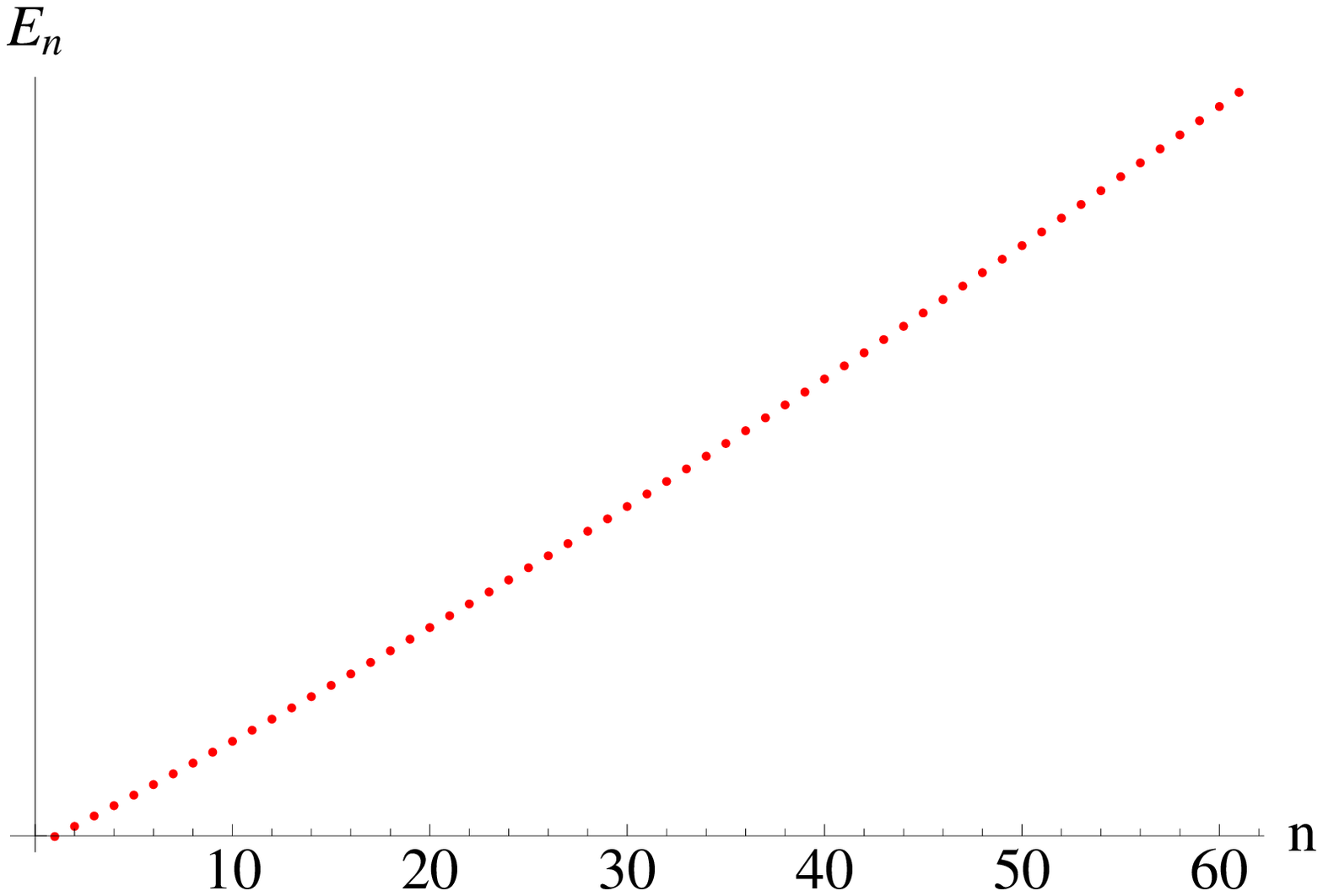} 
\includegraphics[height= 3cm]{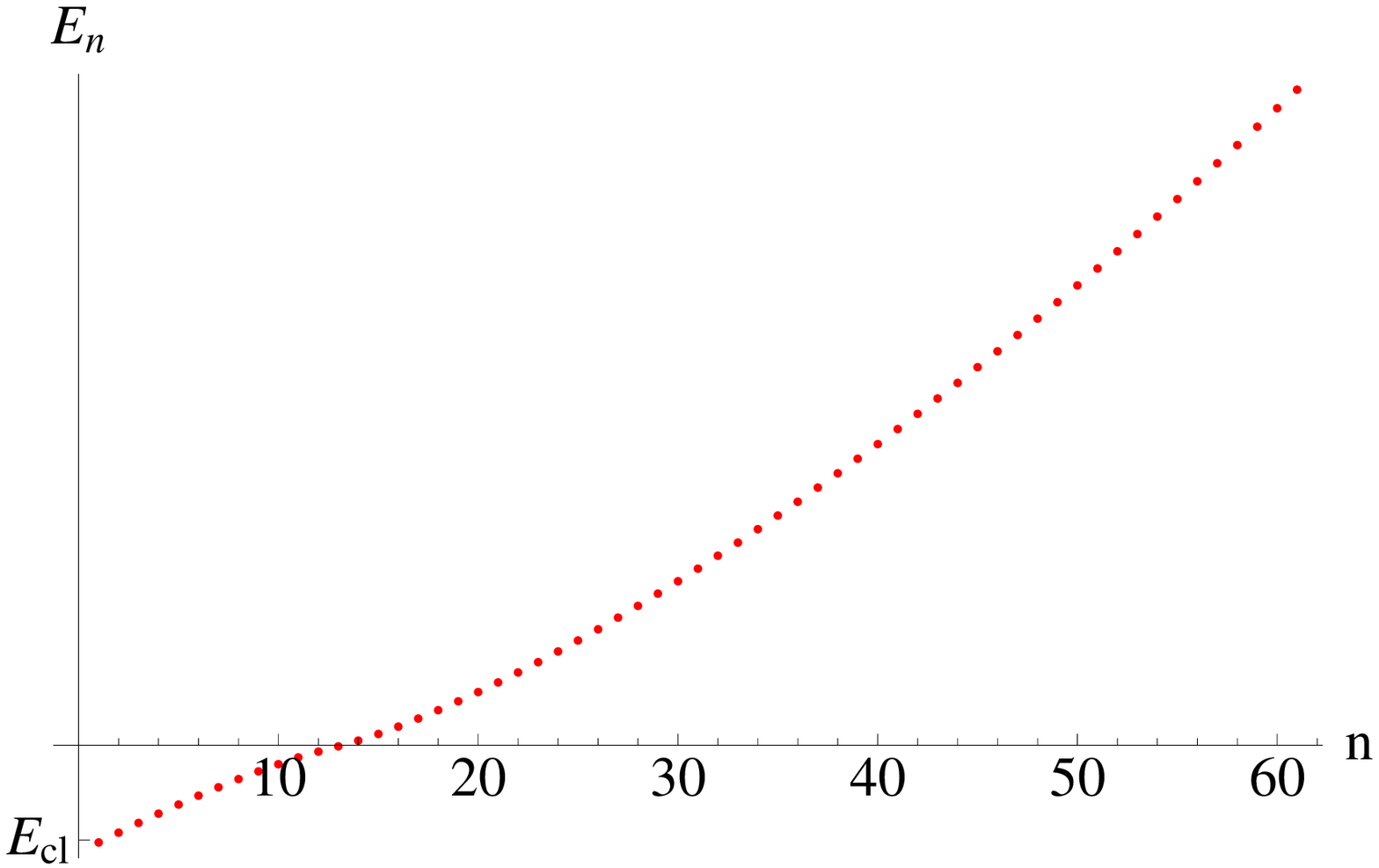} 
\caption{The eigenvalues  of the reduced $H_0$ in the stable regime on the left and in the unstable case on the right. The origin of the Energy axis is the classical critical energy $E_c=2\hbar s \kappa$. 
In the unstable case the ground state energy lies below this line and close to the classical value given by Eq.~(\ref{ground0clas}).
( $\hbar^{-1} = s=30$,  $\kappa= -2\sqrt{2s_{cl}}$ (stable case),  $\kappa= -0.5\sqrt{2s_{cl}}$ (unstable case)). }
\label{figeigen1}
\end{center}
\nonumber
\end{figure}

The eigenvector  equation $H_0 \Psi= E \Psi$ reads:
\begin{equation}
 \hbar^{3\over 2} (n+1)\sqrt{(2s-n)} p_{n+1}  + \hbar^{3\over 2}n \sqrt{(2s+1-n)} p_{n-1} 
 +( 2\hbar \kappa (s-n) -E) p_n =0
 \label{EigEq}
\end{equation}
The eigenvalues are  shown in Figure~\ref{figeigen1} in the stable case and
in the unstable case.

\section{Bohr-Sommerfeld quantization}
\label{semiclassique}
In this section we examine the standard semi-classical analysis of the one spin system. There are two complications as compared to the usual formulae. One comes from the fact that the phase space of the spin degrees of freedom is a 2-sphere and hence has a non trivial topology. The second one is related to the fact that the Schr\"odinger equation is a difference equation and its symbol has a subprincipal part. Let us recall the Schr\"odinger equation:
\begin{equation}
(x+\hbar)\sqrt{ 2s_{cl} -x}\; p(x+\hbar) +  x\sqrt{2s_{cl} +\hbar -x}\; p(x-\hbar) -2 \kappa x \; p(x)= \hbar \epsilon p(x)
\label{schroepsilon1}
\end{equation}
where
$$
\hbar \epsilon = E-2\hbar \kappa s, \quad x=\hbar n
$$
The first thing to do is to compute the Weyl symbol $h(x,\theta)$ of the  Hamiltonian operator. 
In standard notations, it is defined by:
$$
(Hp)(x) = {1\over 2\pi \hbar} \int e^{-{i\over \hbar} (x-y) \theta } h\left({x+y\over 2},\theta \right) p(y) d\theta dy
$$
It is straightforward to check that, for the Hamiltonian of eq.(\ref{schroepsilon1})
$$
h(x,\theta)= 2 \left(x+{\hbar\over 2}\right) \sqrt{2s_{cl} +{\hbar \over 2} -x} \; \cos \theta + 2 \kappa (s_{cl} -x)
$$
Expanding in $\hbar$ we find the principal and sub-principal symbols
$$
h(x,\theta) = h^{0}(x,\theta) + \hbar h^{1}(x,\theta) + \cdots
$$
where:
\begin{eqnarray*}
h^{0}(x,\theta) &=& 2 x \sqrt{2s_{cl}  -x} \; \cos \theta + 2 \kappa (s_{cl} -x) \\
h^{1}(x,\theta) &=& \left( \sqrt{2s_{cl} -x} + {x\over 2 \sqrt{2s_{cl} -x} } \right) \cos \theta
\end{eqnarray*}
We recognize, of course, that $h^{0}(x,\theta)$ is the classical Hamiltonian of the system Eq.(\ref{H0reducedclassical}).
In particular the symplectic form reads:
$$
\omega = dx \wedge d\theta
$$

Note that the definition of the Weyl symbol we have used here is motivated by the simpler
case where the classical phase space can be viewed as the cotangent bundle of a smooth manifold.
In our case, the $\theta$ coordinate is $2\pi$ periodic, and we are not strictly speaking dealing with
the cotangent bundle over the line parametrized by the $x$ coordinate. 
To justify this procedure, we may first imbed our system in a larger phase space, where both
$\theta$ and $x$ run from $-\infty$ to $\infty$. 
Because the symbol $h(x,\theta)$ is $2\pi$ periodic in $\theta$,
the unitary operator $\mathcal{T}$ associated to the $2\pi$ translation of $\theta$ commutes with
the quantum Hamiltonian $H$ associated to the symbol $h(x,\theta)$.  It is then possible to perform
the semi-classical analysis in the enlarged Hilbert space and to project afterwards the states thus obtained
on the subspace of $2\pi$ periodic wave-functions, which are eigenvectors of $\mathcal{T}$ with the
eigenvalue 1. Imposing the periodicity in $\theta$ forces $x$ to be an integer multiple of $\hbar$.
From Eq.~(\ref{schroepsilon1}), we see that the physical subspace, spanned by state vectors $|x=n\hbar\rangle$
with $0 \leq n \leq 2s$, is stable under the action of $H$.  

An alternative approach would be to use a quantization scheme, such as Berezin-Toeplitz 
quantization~\cite{Berezin75}, which allows to work directly with a compact classical phase-space 
such as the sphere. When $H_1$ takes its critical value corresponding to the unstable point, 
it is easy to show that the eigenvector equation~(\ref{EigEq})
can be cast into the Schr\"odinger equation for a pure spin $s$ Hamiltonian given by:
\begin{equation}
H_{\mathrm{eff}}=2\kappa s^{z}+s^{+}\sqrt{s_{cl}-s_{z}}+\sqrt{s_{cl}-s_{z}}\,s^{-}
\end{equation}
Starting from the known Toeplitz symbols of the basic spin operators, one may derive Bohr-Sommerfeld
quantization rules directly from this Hamiltonian $H_{\mathrm{eff}}$~\cite{Charles06}, but we shall not
explore this further in this paper.

Returning to our main discussion we now choose a 1-form $\gamma$ such that: 
\begin{equation}
\gamma (X_{h^0}) = - h^1(x,\theta)
\label{subprincipalform}
\end{equation}
where $X_{h^0}$ is the Hamiltonian vector field associated to the Hamiltonian $h^0(x,\theta)$, which is tangent to the variety $h^0(x,\theta) = E$. Using $x$ as the coordinate on this manifold, we find:
$$
X_{h^0} = 2x \sqrt{2s_{cl}-x} \sin \theta \; \partial_x
$$
Hence, we may choose:
$$
\gamma = -{1\over 2} \left( {1\over x} + {1\over 2(2s_{cl}-x)}\right) \cot \theta \; dx
$$
Under these circumstances, the Bohr-Sommerfeld quantization conditions involve the form $\gamma$ and read
(see e.g. \cite{San99})
\begin{equation}
\Phi_{Reg}(\epsilon) = {1\over 2\pi  \hbar }\int_{C(E)} \alpha + {1\over 2\pi}\int_{C(E)} \gamma + \mu_{C(E)} {1\over 4} =  n
\label{phireg}
\end{equation}
where $C(E)$ is the classical trajectory of energy $E$, $\alpha$ is the canonical 1-form $\omega = d\alpha$,
$ \mu_{C(E)}$ is the Maslov index of this trajectory and $n$ is an integer. Note that the integral of $\gamma$
over $C(E)$ is completely specified by the constraint Eq.(\ref{subprincipalform}).

We now come to the fact that the 2-sphere  has a non trivial topology.  
A consequence is that the canonical 1-form $\alpha$ does not exist globally.  
In the coordinates $x,\theta$ we may choose 
$$
\alpha = x\: d\theta 
$$
but this form is singular in the vicinity of the south pole where $x=2s_{cl}$.
Let us consider a closed path on the sphere parametrized by the segment at constant $x$, 
$\theta$ running from 0 to $2\pi$. The integral of $\alpha$ around this path is $2\pi x$.
When $x$ goes to $2s_{cl}$, the integral goes to $4\pi s_{cl}$ which is the total area of the
sphere. But the limit path for $x=2s_{cl}$, corresponds on the sphere to a trivial loop, fixed 
at the south pole. The consistency of Bohr-Sommerfeld quantization requires then $4\pi s_{cl}$
to be an integer multiple of $2\pi\hbar$. Hence we recover the quantization  condition of the spin:
$2 s$ should be an integer. 

\begin{figure}[hbtp]
\begin{center}
\includegraphics[height= 7cm]{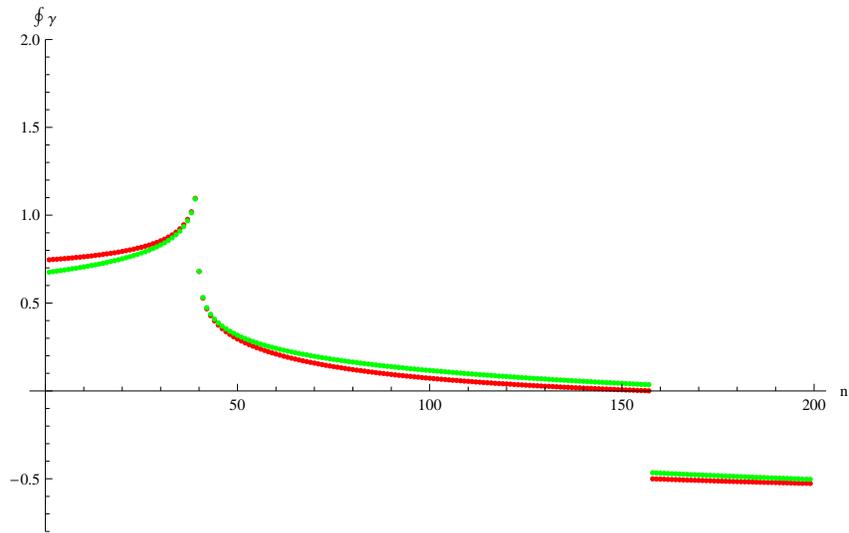}  
\caption{The integral of the sub-dominant symbol as a function of energy (the red curve). 
We see the logarithmic divergence for the energy corresponding to the singular critical point, 
and the gap $-\pi$ at the energy corresponding to the trajectory passing through the south pole. 
The green curve is the asymptotic formula Eq.(\ref{asympt}) in which we have added the gap at the south pole 
to ease comparison and to show that it remains quite accurate even well beyond its expected range of validity.}
\label{figsousdominant}
\end{center}
\nonumber
\end{figure}

Before checking the validity of Eq.(\ref{phireg}) in our system, it is quite instructive to
plot the integral of $\gamma$ along the curve $C(E)$ as a function of $E$. This is shown on Fig.~\ref{figsousdominant}
Here, we see two singularities at the values of $E$ where $C(E)$ goes through
either the north or the south pole. At the south pole $\int_{C(E)} \gamma$ merely jumps by $\pi$,
whereas at the north pole, a jump of $\pi$ is superimposed onto a logarithmic divergence.
We can easily find an asymptotic formula for this integral for energies close to the critical value $E_c= 2 s_{cl} \kappa$:
\begin{equation}
\oint \gamma = {\kappa \over \Omega} \log{ \sqrt{2s_{cl}} | \Delta E | \over 16 \Omega^4} 
+ 2(\pi - \nu) - \pi\; \Theta(\Delta E) + O(\Delta E, \Delta E \log |\Delta E|)
\label{asympt}
\end{equation}
here $\Delta E = E-E_c$, and $\Theta(\Delta E) = 1$ if $\Delta E >0$ and $0$ if $\Delta E<0$. 

\begin{figure}[hbtp]
\begin{center}
\includegraphics[height= 7cm]{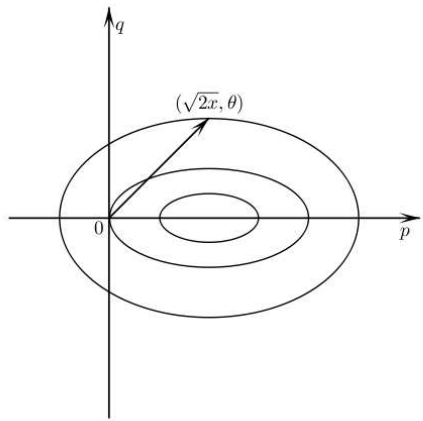}  
\caption{On this figure we see that $-{1\over 2} \oint d\theta$ equals zero if the integral is taken over the small ellipse, while it is equal to $-\pi$ if it is taken over the big ellipse enclosing the origin. This explains why the integral of the subprincipal symbol in polar
coordinates jumps by $\pm \pi$ when the orbit crosses the origin of the polar coordinates.}
\label{ellipses}
\end{center}
\nonumber
\end{figure}

We wish now to show that these jumps are merely an effect of working with polar coordinates
$(x,\theta)$. It is instructive to consider first a quantum system with one degree of freedom,
whose classical phase-space is the $(q,p)$ plane. From any smooth function $h(p,q)$, we define
the quantum operator $\mathcal{O}_{W,h}$ by: 
$$
(\mathcal{O}_{W,h}(\Psi))(q) = {1\over 2\pi \hbar} \int e^{{i\over \hbar} (q-q')p} h\left(p,{q+q'\over 2}\right) \Psi(q') dp\:dq'
$$
Polar coordinates are introduced, at the classical level by:
$$
(p,q)=\sqrt{2x}(\cos \theta,\sin \theta)
$$ 
This definition implies that $dp \wedge dq = dx \wedge d\theta$.
At the quantum level, we start with the usual $(\hat{p}, \hat{q})$ operators.
In the Hilbert space $L^{2}(q)$, we introduce the eigenvector basis $\{|n\rangle\}$ of $\hat{p}^{2}+\hat{q}^{2}$,
such that $(\hat{p}^{2}+\hat{q}^{2}) |n\rangle=\hbar (2n+1)|n\rangle$. We may then set $\hat{x}=\sum_{n\geq 0}\hbar n|n\rangle\langle n|$.
To define $\hat{\theta}$, such that $[\hat{\theta},\hat{x}]=i\hbar$, it is natural to enlarge the Hilbert space,
allowing $n$ to take also negative integer values as well. Then we set $\exp(i\hat{\theta})|n\rangle=|n+1\rangle$.
From $a|n\rangle=\sqrt{n}|n-1\rangle$ and $a^{+}|n\rangle=\sqrt{n+1}|n+1\rangle$, we see that, in the physical
subspace $n \geq 0$, we have $a=\hbar^{-1/2}\exp(-i\hat{\theta})\hat{x}^{1/2}$ and $a^{+}=\hbar^{-1/2}\hat{x}^{1/2}\exp(i\hat{\theta})$.
The Weyl symbols, in the $(x,\theta)$ variables, of these operators are $\hbar^{-1/2}(x+\hbar/2)^{1/2}\exp(\mp i\theta)$.
Using $\hat{p}=(\hbar/2)^{1/2}(a+a^{+})$ and $\hat{q}=(\hbar/2)^{1/2}i(a-a^{+})$, we deduce that the Weyl symbols of
$\hat{p}$ and $\hat{q}$ in the $(x,\theta)$ variables are respectively $\sqrt{2x+\hbar}\:\cos\theta$
and $\sqrt{2x+\hbar}\:\sin\theta$. So the Weyl symbol is not invariant under the non-linear change of coordinates from
$(p,q)$ to $(x,\theta)$. For linear functions of $p$ and $q$ the symbol in $(x,\theta)$, denoted by $h_{\mathrm{pol}}(x,\theta)$ 
is obtained from $h(p,q)$ by substituting $p$ and $q$ by their classical expressions as functions of $(x,\theta)$
and then by shifting $x$ into $x+\hbar/2$. Such a simple rule does not hold for more complicated functions $h(p,q)$.
Nevertheless, one can show that the substitution of $x \rightarrow x+\hbar/2$ gives the symbol $h_{\mathrm{pol}}(x,\theta)$
up to first order in $\hbar$. Practically, this means that:
$$
h_{\mathrm{pol}}(x,\theta)=h(\sqrt{2x+\hbar}\cos\theta,\sqrt{2x+\hbar}\sin\theta) \;\;\;\mathrm{mod}\;\mathcal{O}(\hbar^{2})
$$
or, more explicitely:
$$
h_{\mathrm{pol}}(x,\theta)=h(p,q)+\frac{\hbar}{4x}\left(p\frac{\partial h}{\partial p}(p,q)+q\frac{\partial h}{\partial q}(p,q)\right)
+\mathcal{O}(\hbar^{2})
$$
where $(p,q)=\sqrt{2x}(\cos \theta,\sin \theta)$.
This has the following important consequence: even if $h(p,q)$ has no subprincipal part (i.e. it does not depend on
$\hbar$), $h_{\mathrm{pol}}(x,\theta)$ does acquire a subprincipal part $h_{\mathrm{pol}}^{1}(x,\theta)$. 
Now, along any classical trajectory associated to
$h(p,q)$, we have:
$$
p\frac{\partial h}{\partial p}(p,q)+q\frac{\partial h}{\partial q}(p,q)=p\dot{q}-q\dot{p}=2x\dot{\theta}, 
$$
so $h_{\mathrm{pol}}^{1}(x,\theta)=\dot{\theta}/2$. This striking result implies that, when we work in polar
coordinates, the $\gamma$ form associated to $h_{\mathrm{pol}}^{1}(x,\theta)$ may be chosen as $-d\theta /2$. This
immediately explains why $\int_{C(E)} \gamma$ jumps by $-\pi$ when the $h(p,q)=E$ orbit crosses the origin of the
polar coordinates, because then $\int_{C(E)} d\theta$ jumps from 0 to $2\pi$, see Fig.~\ref{ellipses}.
Note that, in general, we expect
$h(p,q)$ to have also a subprincipal part $h^{1}(p,q)$. But this gives an additional term to $h_{\mathrm{pol}}^{1}(x,\theta)$
which is deduced from $h^{1}(p,q)$ simply by the classical change of variables from $(p,q)$ to $(x,\theta)$. This term has
no reason to display any singularity when the classical trajectory goes through the origin. Remark that around the south pole, if we set
$$
x=2s_{cl}-y, \quad p=\sqrt{2y}\cos \theta, \quad q = \sqrt{2y}\sin \theta
$$
then we can expand
$$
h_0 = -2 s_{cl}( \kappa +s_{cl}) + \kappa \left( \left(p+{\sqrt{2}  s_{cl}\over \kappa} \right)^2 + q^2 \right)
$$
and we see that our system is equivalent to a shifted harmonic oscillator.

Note that on the sphere, polar coordinates have two singularities, at the north and at the south poles, which
correspond to $x=0$ and $x=2s_{cl}$ respectively. To generalize the above analysis to the sphere, we may thus use two
charts $(p,q)$ and $(p',q')$ such that:
\begin{eqnarray*}
(p,q) & = & \sqrt{2x}(\cos \theta,\sin \theta) \\
(p',q') & = & \sqrt{2x'}(\cos \theta',\sin \theta') \\
x+x' & = & 2s_{cl} \\
\theta + \theta' & = & 0
\end{eqnarray*}
We therefore get $dp \wedge dq = dx \wedge d\theta = dx' \wedge d\theta' = dp' \wedge dq'$. It is 
interesting to note that the Weyl symbols $h_{\mathrm{pol}}(x,\theta)$ and $h'_{\mathrm{pol}}(x',\theta')$
are obtained from each other by the classical transformation from $(x,\theta)$ to $(x',\theta')$.
On the other hand, $h(p,q)$ and $h'(p',q')$ corresponding to the same quantum Hamiltonian in the physical
subspace $0\leq x=2s_{cl}-x' \leq 2s_{cl}$ do not coincide, because shifting $x$ into $x+\hbar/2$ amounts to
shifting $x'$ into $x'-\hbar/2$ instead of $x'+\hbar/2$. This implies that it is impossible to construct
a quantum operator in the $2s+1$ dimensional Hilbert space of a spin $s$ for which both subprincipal
parts of $h(p,q)$ and $h'(p',q')$ would vanish.

To formulate the Bohr-Sommerfeld rule, it is convenient to use the $(p,q)$ coordinates when we consider
a set of classical orbits which remain at a finite distance from the south pole. In these coordinates, the
subleading term $\int_{C(E)} \gamma$ and the Maslov index are both continuous when $C(E)$ crosses the north pole.
Going back to $(x,\theta)$ coordinates, we see that the jump in $\int_{C(E)} \gamma$ has to be compensated by a
jump in the Maslov index. So the Bohr-Sommerfeld formula~(\ref{phireg}) can be expressed in $(x,\theta)$
coordinates, provided these jumps in the Maslov index are taken into account.

To check Eq.(\ref{phireg}) we evaluate $\Phi_{Reg}(\epsilon_n)$ for the {\em exact} values $\epsilon_n$ 
and we define the defect $\delta_n$ by:
 $$
 \Phi_{Reg}(\epsilon_n) = n+ \delta_n
 $$
 Denoting
 $$
 I_{-1}= {1\over 2\pi  \hbar }\int_{C(E)} \alpha, \quad I_0 = {1\over 2\pi}\int_{C(E)} \gamma
 $$
 \begin{eqnarray*}
 \delta_n = & I_{-1}+I_0  - (n+{1\over 2}),&\quad {\rm if} \;\;\epsilon_n \le 2\hbar \kappa s \\
 \delta_n = &  I_{-1}+I_0  + {1\over 2} - (n+{1\over 2}),&\quad  {\rm if}\;\; 2\hbar \kappa s \le \epsilon_n \le - 2\hbar \kappa s  \\
 \delta_n = &  I_{-1}+I_0  + 1- (n+{1\over 2}) ,& \quad   {\rm if} \;\; \epsilon_n \ge - 2\hbar \kappa s  
\end{eqnarray*}

\begin{figure}[hbtp]
\begin{center}
\includegraphics[height= 6cm]{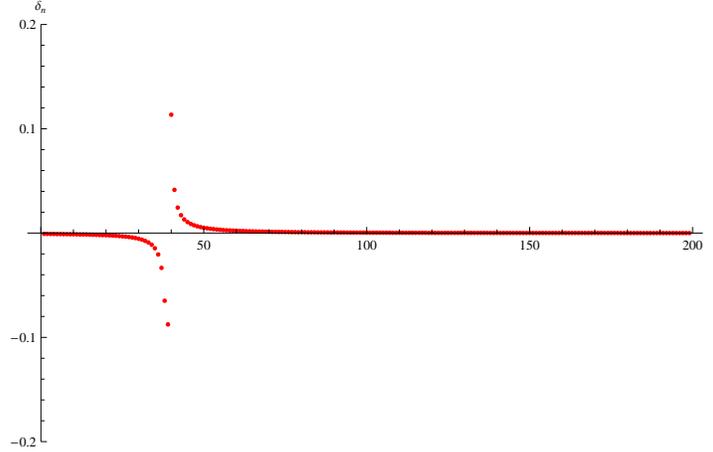} 
\caption{To test the quality of the Bohr-Sommerfeld quantization, we plot 
the default $\delta_n$ as a function of $n$ which labels the energy eigenstates. 
We see that the accuracy of the usual Bohr-Sommerfeld rule is very good
away from the critical energy. In particular there is nothing special at the south pole.}
\label{deltan}
\end{center}
\nonumber
\end{figure}

The excellent accuracy of this Bohr-Sommerfeld rule is clearly visible on Fig.~\ref{deltan}, 
which shows the defect $\delta_n$ as a function of $n$. We see that it remains small everywhere, 
and we emphasize that it does not exhibit any accident when $\epsilon_n$ crosses the value $-2\kappa s_{cl}$, 
corresponding to the south pole. On the other hand, something special happens around $\epsilon_n = 2\kappa s_{cl}$
(the energy of the north pole), which cannot be attributed to the use of polar coordinates, 
but which reflects the crossing through the singular orbit associated to the pinched torus.
A detailed description of the energy spectrum in the vicinity of the classical unstable point is the
goal of next section.

\section{Generalized Bohr-Sommerfeld rule}
\label{semiclassiquesing}
This region of the spectrum requires a different quantization 
rule~\cite{Colin94a,Colin94b,Brummelhuis95,Child98,Colin99,San00,Keeling09} because the classical motion near the turning 
point at $x_{min}$ is strongly affected by the presence of the unstable point at $x=0$. Therefore, we first need
to analyze the small $x$ behavior of the energy eigenstates, when $E$ is close to $2\kappa s_{cl}$.
In the intermediate regime, between $x_{min}(E)$ and $x_{max}(E)$, a standard WKB analysis is quite acurate.
Near $x_{max}(E)$, we have a regular turning point, which can be described, as usual, by an Airy function.
Gluing the wave function of the eigenstates between these three different regimes 
will give us the generalized Bohr Sommerfeld quantization rules.

\subsection{Small $x$ analysis}

Here, we shall recast the normal form around the singular point, studied in 
section~\ref{secclassicalmonodromy} within the framework of the reduced system. For this,
we assume $n<< 2s$ or $x=\hbar n << 2s_{cl}$.  In that approximation the Schr\"odinger equation  
Eq.(\ref{EigEq}) becomes:
\begin{equation}
E p_n= \hbar \sqrt{2s \hbar} (n+1) p_{n+1}  + \hbar  \sqrt{2s\hbar } \;n p_{n-1} 
 + 2\hbar \kappa (s-n)  p_n 
 \label{eigschroerduite2}
\end{equation}
This equation is linear in $n$ and can be solved by Laplace-Fourier transform. Letting:
\begin{equation}
\Psi^{\rm Small}(\theta)= \sum_{n=0}^\infty p_n e^{in \theta}
\label{Psidef}
\end{equation}
the equation  reads:
$$
H_0 \Psi^{\rm Small} = E  \Psi^{\rm Small}, \quad E = E_c +  \hbar \epsilon
$$
with: 
\begin{equation}
H_0 =-i\hbar ( \sqrt{2s_{cl}}(e^{i\theta} + e^{-i\theta} ) -2\kappa) \partial_{\theta}
+\hbar (2 \kappa s + \sqrt{2s_{cl} }e^{i\theta} )
\label{H0theta}
\end{equation}
Let us set:
$$
z =  e^{i\theta},\quad {d\over d \theta} = i z {d\over d z}
$$
the equation becomes:
$$
\left[ (\sqrt{2s_{cl}} (z + z^{-1}) -2\kappa ) z{d\over dz} + \sqrt{2s_{cl}} z \right] \Psi^{\rm Small} = \epsilon \Psi^{\rm Small}
$$
We are looking for a solution which is analytic in a neighborhood of the origin in the
$z$-plane. This can be written as:
$$
\Psi^{\rm Small}(z, \epsilon) = \left(1-\frac{z}{z_+}\right)^{\Delta_+} \left(1-\frac{z}{z_-}\right)^{\Delta_-}
$$
where $z_\pm$ are the solutions of the second order equation already introduced before:
$$
z^2 -2{\kappa \over \sqrt{2s_{cl}}} z +1 = 0, \quad z_\pm = {\kappa \pm i\Omega \over \sqrt{2s_{cl}}}
= -e^{\mp i \nu}
$$
and
$$
\Delta_\pm = -{1\over 2 } \mp i{\epsilon - \kappa\over 2 \Omega} \equiv -{1\over 2 } \mp if(E)
$$
Note that, when $\hbar$ is small, the leading term in $f(E)$ is $(E-E_c)/2\hbar \Omega$.
It is quite instructive to express directly this wave-function in the $\theta$ variable. Denoting
$$
\chi(\theta) = (\cos\theta+\cos\nu)^{-1/2}\exp i\left[f(E)
\log\left(\frac{|\cos\frac{\theta-\nu}{2}|}{|\cos\frac{\theta+\nu}{2}|}\right)-\frac{\theta}{2}\right] 
$$
we have
\begin{eqnarray*}
\Psi^{\rm Small}(\theta)  = &\quad  \chi(\theta),            &  \;\;\;-\pi+\nu \leq \theta \leq \pi-\nu \\
\Psi^{\rm Small}(\theta)  = & i\exp(-\pi f(E))\chi(\theta),&\;\;\; \pi-\nu \leq \theta \leq \pi  \\
\Psi^{\rm Small}(\theta)  = & -i\exp(-\pi f(E))\chi(\theta),& \;\;\; -\pi \leq \theta \leq -\pi + \nu 
\end{eqnarray*}
Notice that this  has precisely the expected form in the semi-classical limit $\hbar \rightarrow 0$.
Indeed, when $E>E_c$, the wave function is mostly confined to
the interval $-\pi+\nu \leq \theta \leq \pi-\nu$, and it is exponentially small in the classically
forbidden regions $\pi-\nu \leq \theta \leq \pi$ and $-\pi \leq \theta \leq -\pi + \nu$.
Such exponentially small prefactor is reminiscent of the behavior of a tunneling amplitude. This is
consistent with our treatment of $\theta$ as an unbounded variable, subjected to a $2\pi$-periodic 
Hamiltonian. Enforcing integer values of $n$ requires the wave-function $\Psi^{\rm Small}(\theta)$
to be $2\pi$-periodic in $\theta$. If $\Psi^{\rm Small}(\theta)$ were identically zero in the
classically forbidden intervals, its Fourier transform would develop a tail for negative values
of $n$. So these evanescent parts of the wave-function are required in order to ensure that the
wave function belongs to the physical Hilbert space $n \geq 0$. 
Note that when $E<E_c$, the classically forbidden region becomes the intervals  
$-\pi \leq \theta \leq -\pi + \nu$ and $\pi-\nu \leq \theta \leq \pi$.

In the classically allowed regions, the semi-classical wave function is expected to take the
textbook form:
\begin{equation}
\Psi^{\rm Small}(\theta)=a(\theta)\exp i \left(\frac{S_{0}(\theta)}{\hbar}+S_{1}(\theta)\right)
\end{equation}
where the three functions $a$, $S_{0}$ and $S_{1}$ take real values.
These functions satisfy the standard equations~\cite{Duistermaat74}, 
involving the principal and subprincipal
symbols $h^{0}(x,\theta)=E_c+2\sqrt{2s_{cl}}x(\cos\theta+\cos\nu)$ and 
$h^{1}(x,\theta)= \sqrt{2s_{cl}}\cos\theta$:
\begin{eqnarray}
h^{0}(S'_{0}(\theta),\theta) & = & E \label{fonctiondephase} \\
\frac{d}{d\theta}\left(a^{2}(\theta)\partial_x h^{0}(S'_{0}(\theta),\theta)\right) & = & 0 
\label{transportamplitude} \\
\partial_x h^{0}(S'_{0}(\theta),\theta)S'_{1}(\theta) & = & - h^{1}(S'_{0}(\theta),\theta)
\end{eqnarray}
From the above expressions for $\Psi^{\rm Small}(\theta)$, we can check that it has
{\em exactly} this semi-classical form. Most likely, this occurs because
$\Psi^{\rm Small}(\theta)$ is an eigenstate of a
quantum Hamiltonian which is derived by symplectic reduction  
from the quadratic Hamiltonian of the normal form discussed in section~\ref{secclassicalmonodromy}.
Quadratic Hamiltonians are particularly well-behaved with respect to the $\hbar \rightarrow 0$
limit in the sense that the Bohr-Sommerfeld formula gives the exact energy spectrum, and also
that the full quantum propagator obeys similar equations as ~(\ref{fonctiondephase}) and 
(\ref{transportamplitude}).

Let us now discuss the $x=n\hbar$ representation which is quite useful from the physical standpoint.
$$
p^{\rm Small}(x,\epsilon) = \oint_{C_0} {dz \over 2i\pi }z^{-{x\over \hbar} - 1} \Psi(z)=  \oint_{C_0} {dz \over 2i\pi }z^{-{x\over \hbar} - 1} \left(1-\frac{z}{z_+}\right)^{\Delta_+} \left(1-\frac{z}{z_-}\right)^{\Delta_-}
$$
where $C_0$ is a small contour around the origin. 
When $E$ is sufficiently far (in a sense to be precised below) from its critical value $E_c$,
we may use the saddle point approximation to evaluate $\Psi^{\rm Small}(x,\epsilon)$.
This yields again the expected semi-classical form, namely: 
\begin{equation}
p^{\rm Small}(x)=b(x)e^{\pm i\pi/4}\exp -i \left(\frac{W_{0}(x)}{\hbar}+W_{1}(x)\right)
\end{equation}
where $b(x)$, $W_{0}(x)$ and $S_{1}(x)$ satisfy:
\begin{eqnarray}
h^{0}(x,W'_{0}(x)) & = & E  \label{fonctiondephaseW} \\
\frac{d}{dx}\left(b^{2}(x)\partial_{\theta} h^{0}(x,W'_{0}(x))\right) & = & 0 
\label{transportamplitudeW} \\
\partial_{\theta} h^{0}(x,W'_{0}(x))W'_{1}(x) & = & - h^{1}(x,W'_{0}(x))
\label{subleadingphaseW} 
\end{eqnarray}
Note that $W_{0}(x)$ is the Legendre transform of $S_{0}(\theta)$, that is
$S'_{0}(\theta)=x$, $W'_{0}(x)=\theta$ and $S_{0}(\theta)+W_{0}(x)=x\theta$.
It is interesting to write down explicitely this semi classical wave-function
when $x \gg 1$. It reads:
$$
p^{\rm Small}_{\rm sc}(x)=B_{+}(E)\frac{e^{-i(\pi-\nu)x/\hbar}}{(x/\hbar)^{\frac{1}{2}-if(E)}}
+B_{-}(E)\frac{e^{i(\pi-\nu)x/\hbar}}{(x/\hbar)^{\frac{1}{2}+if(E)}}
$$
where $B_{-}(E)=\bar{B}_{+}(E)$ and:
\begin{equation}
\frac{B_{+}(E)}{B_{-}(E)}=e^{i(\nu-\pi/2)}\exp i \left(2f(E)\log\frac{4\hbar\Omega}{|E-E_c|}
+\frac{E-E_c}{\hbar\Omega}\right)
\end{equation}
We note that the phase factor diverges when $E$ reaches $E_c$. This behavior is induced
by the contribution of the subprincipal symbol to the phase of the wave-function.
The prefactor $\exp(i(\nu-\pi/2))$ in the above expression does not jump when $E$
crosses the critical value $E_c$. This is consistent with the analysis of the previous
section, because we do find a $\pi$ jump in the Maslov index that is compensated by a
$\pi$ jump in the contribution of the subprincipal symbol.  

Let us now precise the validity domain of the stationary phase approximation.
When we integrate over $\theta$ to compute the Fourier transform $p^{\rm Small}(x)$,
we get an oscillating integral whose phase factor can be approximated by
$\exp(iS''_{0}(\theta(x))(\theta-\theta(x))^{2}/(2\hbar))$ where $\theta(x)$ is
one of the two saddle points defined implicitely by $S'_{0}(\theta(x))=x$. 
This oscillating gaussian has a typical width $\langle\Delta\theta^{2}\rangle=\hbar/|S''_{0}(\theta(x))|$.
We find that: 
$$
S''_{0}(\theta(x))=\frac{E-E_c}{2\sqrt{2s_{cl}}}\frac{\sin\theta}{(\cos\theta+\cos\nu)^{2}}
$$
When $E$ goes to $E_c$ at fixed $x$, $\theta(x)$ goes to $\pm (\pi-\nu)$, because
$\cos\theta+\cos\nu=E/(2\sqrt{2s_{cl}}x)$. So $|S''_{0}(\theta(x))|\simeq 2\Omega x^{2}/|E-E_{c}|$.
On the other hand, we have seen that the amplitude of $\Psi^{\rm Small}(\theta)$ diverges when
$\theta=\pm (\pi-\nu)$.  The distance $\delta\theta$ between $\theta(x)$ and the closest
singularity goes like $|E-E_{c}|/2\Omega x$. The stationary phase approximation holds
as long as $\theta(x)$ is far enough from the singularities, that is if 
$\langle\Delta\theta^{2}\rangle<<(\delta\theta)^{2}$, or equivalently,
\begin{equation}
|E-E_c| \gg 2\hbar \Omega
\end{equation}
 
When this condition is not fullfilled, the stationary phase approximation breaks down.
The dominant contribution to $p^{\rm Small}(x)$ comes from the vicinity of $\theta=\pm (\pi-\nu)$.
We have therefore to consider a different asymptotic regime, where $\hbar$ goes to $0$ while
$\epsilon=(E-E_{c})/\hbar$ remains fixed. In this case, the two exponents $\Delta_{+}$, $\Delta_{-}$
are fixed, and the only large parameter in the Fourier integral giving $p^{\rm Small}(x)$
is $x/\hbar$. One then obtains:
\begin{equation}
p^{\rm Small}(x)=A_{+}(E)\frac{e^{-i(\pi-\nu)x/\hbar}}{(x/\hbar)^{\frac{1}{2}-if(E)}}
+A_{-}(E)\frac{e^{i(\pi-\nu)x/\hbar}}{(x/\hbar)^{\frac{1}{2}+if(E)}}\equiv p_{+}^{\rm Small}(x)+p_{-}^{\rm Small}(x)
\label{psmallx}
\end{equation}
with $A_{-}(E)=\bar{A}_{+}(E)$ and:
\begin{equation}
\frac{A_{+}(E)}{A_{-}(E)}=e^{i(\nu-\pi/2)}\frac{\Gamma(\frac{1}{2}-if(E))}{\Gamma(\frac{1}{2}+if(E))}
\exp i \left(2f(E)\log (2\sin\nu) \right)
\end{equation}
This analysis first shows that, in spite of the breakdown of the stationary phase approximation,
$p^{\rm Small}(x)$ is still given at large $x$ by a sum of two semi-classical  
wave-functions associated to the same principal and subprincipal symbols $h^{0}$ and $h^{1}$.
This implies that there will be a perfect matching between $p^{\rm Small}(x)$ and the WKB
wave functions built at finite $x$ from the full classical Hamiltonian. This will be discussed
in more detail below. The only modification to the conventional WKB analysis lies in the
phase factor $A_{+}/A_{-}$ between the ongoing and outgoing amplitudes, which differs markedly 
from the semi-classical $B_{+}/B_{-}$. In particular, we see that the full quantum treatment at
fixed $\epsilon$ provides a regularization of the divergence coming from the subprincipal symbol.
Indeed, the singular factor $\log(4\hbar\Omega/|E-E_c|)$ is replaced by the constant $\log(2\sin\nu)$.
Such a phenomenon has been demonstrated before for various models~\cite{Colin99,San00}.

\subsection{WKB analysis}
We return to the Schr\"odinger equation: 
\begin{equation}
(x+\hbar)\sqrt{ 2s_{cl} -x}\; p(x+\hbar) +  x\sqrt{2s_{cl} +\hbar -x}\; p(x-\hbar) -2 \kappa x \; p(x)= \hbar \epsilon p(x)
\label{schroepsilon}
\end{equation}
where:
$$
\hbar \epsilon = E-E_{c}
$$
We try to solve this equation by making the WKB Ansatz:
$$
p^{\rm WKB}(x) = e^{{-i\over \hbar}(W_{0}(x)+\hbar W_{1}(x))} b(x)
$$
Expanding in $\hbar$,  to order $\hbar^0$ we find

\begin{equation}
x \sqrt{2s_{cl}-x} \cos W'_{0} - \kappa x =0
\label{WKB}
\end{equation}
which is nothing but the Hamilton-Jacobi equation on the critical variety. It is
identical to Eq.~(\ref{fonctiondephaseW}) where $h^{0}(x,\theta)$ is the complete principal symbol,
and the energy is taken to be $E_c$. In this procedure, the energy difference $\hbar \epsilon$
is viewed as a perturbation, which can be treated by adding the constant term $-\epsilon$ 
to the complete subprincipal symbol $h^{1}(x,\theta)$.
Alternatively, we may write the Hamilton-Jacobi equation as:
$$
e^{i W'_{0}} = z(x), \quad z^2(x) -{2\kappa \over \sqrt{2s_{cl} -x}} z(x) + 1 =0
$$
The solution  of the quadratic equation is:
\begin{equation}
z_\pm(x) = {1\over \sqrt{2s_{cl}-x}} (\kappa \pm i \sqrt{\Omega^2-x})=
\left( {\kappa \pm i  \sqrt{\Omega^2-x} \over \kappa \mp i  \sqrt{\Omega^2-x}} \right)^{1/2}
\label{zpmx}
\end{equation}
so that $z_\pm(x)\vert_{x=0}  = e^{\pm i(\pi - \nu)}$. Hence:
\begin{equation}
e^{-{i\over \hbar} W_{0}(x)} = e^{\mp {i\over \hbar} \left(\pi x+\kappa \sqrt{\Omega^2-x} + {i\over 2} (2s_{cl}-x) 
\log{ \kappa +i \sqrt{\Omega^2-x} \over \kappa -i \sqrt{\Omega^2-x}  } \right)} 
\label{soluWKB}
\end{equation}
 where the two signs refer to the two branches of the classical trajectory. Notice that they also correspond to the two determinations of the square root $\sqrt{\Omega^2-x}$.

At the next order in $\hbar$ the equation for $b(x)$ is: 
\begin{equation}
\frac{d}{dx}\left(b^{2}(x)x\sqrt{\Omega^{2}-x}\right)=0 
\label{WKBstationary}
\end{equation}
Therefore, we may choose $b(x)=(x\sqrt{\Omega^{2}-x})^{-1/2}$.

The correction $W_{1}(x)$ to the phase function satisfies Eq.~(\ref{subleadingphaseW}) with
$h^{1}(x,\theta)$ replaced by $h^{1}(x,\theta)-\epsilon$, that is:
$$
\partial_{\theta} h^{0}(x,W'_{0}(x))W'_{1}(x) = - h^{1}(x,W'_{0}(x))+\epsilon 
$$ 
$W'_{1}(x)$ is the sum of two terms, $W'_{1}(x)=-\gamma(x)+\epsilon \delta \beta(x)$,
where $\gamma(x)dx$ is simply the one-form $\gamma$ defined by Eq.~(\ref{subprincipalform}),
evaluated on the critical trajectory, and $\delta \beta(x)=\left(\partial_{\theta} h^{0}(x,W'_{0}(x))\right)^{-1}$.
To understand better the meaning of $\delta \beta (x)$, we start with the action integral $\int_{C(E)} \beta$,
where $\beta = \theta dx$ is closely related to the canonical 1-form $\alpha = x d\theta$,
because $\alpha+\beta=d(x\theta)$. Note that, by contrast to $\alpha$, $\beta$ is defined on the
sphere only after choosing a determination of the longitude $\theta$. Let us now study
the variation of this action integral when $E$ moves away from $E_c$ and is changed into 
$E_c + \hbar \epsilon$. We have:
$$
 \int_{C(E_c+ \hbar \epsilon)} \beta - \int_{C(E_{c})} \beta = \hbar \epsilon  \int_{C(E_c)} 
{\partial\theta\over \partial E} dx
$$
But $\theta(x,E)$ satisfies $h^{0}(x,\theta(x,E))=E$, so:
$$
\partial_{\theta}h^{0}(x,\theta(x,E))\frac{\partial \theta}{\partial E}(x,E)=1
$$ 
and finally:
$$
\int_{C(E_c+ \hbar \epsilon)} \beta - \int_{C(E_{c})} \beta = \hbar \epsilon  \int_{C(E_c)} \delta \beta(x) dx
$$

Explicitely, the equation for $W_{1}(x)$ reads:
\begin{equation}
W'_{1}(x) = {\pm 1\over 2\sqrt{\Omega^2-x}} 
\left( {\kappa \over 2(2s_{cl} -x)} +  {\kappa - \epsilon \over x} \right) 
\label{WKBstationary2}
\end{equation}
So we have:
\begin{equation}
e^{-iW_{1}(x)} = \exp\left(\pm {1\over 4} \log{ \kappa +i \sqrt{\Omega^2-x} \over
 \kappa -i \sqrt{\Omega^2-x}  } \mp i {\epsilon -\kappa\over 2 \Omega} \log{ \Omega + \sqrt{\Omega^2-x} \over \Omega - \sqrt{\Omega^2-x}}\right) 
\label{expressionW1}
\end{equation}

We can now expand $p^{\rm WKB}(x)$ when $x$ is small. We find:
\begin{equation}
p_\pm^{\rm WKB}(x) \simeq A_\pm^{\rm WKB} (\epsilon)\frac{e^{\mp i(\pi-\nu){x\over \hbar}}}{(x/\hbar)^{\frac{1}{2}\mp if(E)}}
\label{wkbsmall}
\end{equation}
where:
$$
A_\pm^{\rm WKB} (\epsilon)= {1\over \sqrt{\Omega}} e^{\mp i \left[ { \kappa \Omega \over \hbar }
+(2s+\frac{1}{2})\nu+{\epsilon-\kappa \over 2\Omega}\log \left( {4 \Omega^2 \over \hbar} \right)\right]}
$$
Notice that the phase factor is the sum of two terms which have simple geometrical interpretations:
$$
S^+_{cl} =   \kappa \Omega + 2s_{cl}\nu  = \pi \Omega^{2}-\int_0^{\Omega^2}  \theta(x)dx
= \int_{C_+} \alpha
$$
is the classical action computed on the upper half of the classical trajectory, 
and
$$
\frac{\nu}{2} +  {\epsilon-\kappa \over 2\Omega}\log \left( {4 \Omega^2 \over \hbar} \right) =
\int^{\Omega^2}_0\hskip-0.7cm  \backprime \hskip .5cm \tilde\gamma(x) dx
$$
is the regularized integral of the subprincipal symbol $\tilde\gamma(x)=\gamma(x)-\epsilon\delta\beta (x)$ 
on the same trajectory:
$$
\int^{\Omega^2}_0\hskip-.7cm  \backprime \hskip .55cm \tilde\gamma(x) dx = \lim_{x_A\to 0}
\left(\int_{x_A}^{\Omega^{2}} \tilde\gamma(x) dx + f(E) \log { x_{A} \over \hbar } \right)
$$

Similarly, setting  $ x = \Omega - |\xi |$ and expanding in $|\xi|\over \Omega$ and $|\xi|\over \kappa$ we find to leading order:
\begin{equation}
p^{\rm WKB}_\pm(x) \simeq {1\over \Omega | \xi|^{1/4}} e^{\pm{2i \over 3 \hbar \kappa} |\xi|^{3/2}
\pm i\left( {1\over 2\kappa} - {\epsilon-\kappa \over \Omega^2}\right) |\xi|^{1/2}
}
\label{wkbairy}
\end{equation}
The relevant wave function is of course a linear combination of $p_\pm^{\rm WKB}(x)$.

\subsection{Airy function analysis}

Again, we start with the Schr\"odinger equation Eq.~(\ref{schroepsilon}). We set:
$$
p^{\rm Airy}(x) = (-1)^n \tilde{p}^{\rm Airy}(x),\quad x = n\hbar
$$
\begin{equation}
(x+\hbar)\sqrt{ 2s_{cl} -x}\;  \tilde{p}^{\rm Airy}(x+\hbar) +  x\sqrt{2s_{cl} +\hbar -x}\;  \tilde{p}^{\rm Airy}(x-\hbar) +2 \kappa x  \tilde{p}^{\rm Airy}(x)= -\hbar \epsilon \tilde{p}^{\rm Airy}(x)
\end{equation}
and we expand $x=\Omega^2 + \xi$, keeping the terms linear in $\xi$. Remembering that 
$\sqrt{\kappa^2} = -\kappa$, we find:
$$
{d^2\over d\xi^2} \tilde{p}^{\rm Airy}+ {1\over \Omega^2}\left(1-{\Omega^2\over 2\kappa^2}\right)   
{d\over d\xi}\tilde{p}^{\rm Airy} -{1\over \hbar^2 \kappa \Omega^2}
\left( \hbar \epsilon - {\hbar \Omega^2\over 2\kappa} - \hbar \kappa  - {\Omega^2\over \kappa} \xi \right)
 \tilde{p}^{\rm Airy}=0
$$
The unique solution which decreases exponentially  in the classically forbiden region $\xi >0$ is proportional to the Airy function:
$$
\tilde{p}^{\rm Airy}(\xi) = e^{-c\; \xi } 
{\rm Airy}( a^{1/3}(\xi- b ) )
$$
where:
$$\ a= {1\over \hbar^2 \kappa^2},\quad 
b=\hbar\left({1\over 2} -  { \kappa (\epsilon-\kappa)\over \Omega^2}\right), \quad
c= {1\over 2 \Omega^2} ( 1-{\Omega^2\over 2\kappa^2}) 
$$
When $\xi <0$ it  behaves like:
$$
{\sin ( {2\over 3} |X|^{2/3} + {\pi\over 4} ) \over |X|^{1/4}}, \quad X =  a^{1/3}(\xi- b )
$$
We expand in $|\xi|\over \kappa$ and $|\xi|\over \Omega$. One has:
$$
{2\over 3}|X|^{3/2} =- {2\over 3 \hbar \kappa} |\xi|^{3/2} -  \left(
{1\over 2 \kappa}-{ (\epsilon-\kappa)\over \Omega^2} \right) |\xi|^{1/2}
$$
In that approximation  $e^{-c\; \xi } \simeq 1$, and we get:
\begin{equation}
\tilde{p}^{\rm Airy}(\xi) = C  { \sin \left({2\over 3 \hbar \kappa} |\xi|^{3/2} +  \left(
{1\over 2 \kappa}-{ (\epsilon-\kappa)\over \Omega^2} \right) |\xi|^{1/2}- {\pi\over 4}\right)\over \sqrt{\pi} |\xi|^{1/4}}
\label{airy}
\end{equation}

\subsection{Gluing the parts together}

\begin{figure}[hbtp]
\begin{center}
\includegraphics[height= 5cm,width=15cm]{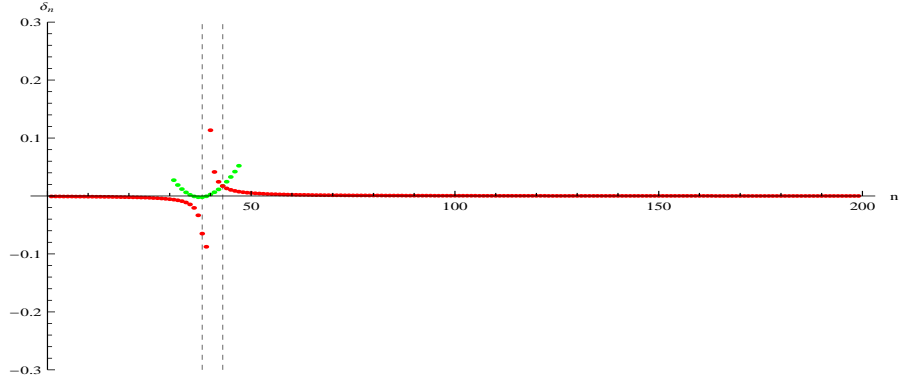} 
\caption{To compare the usual and the singular Bohr-Sommerfeld rules, the default $\delta_n$ is plotted as a function of $n$. 
The red dots are obtained by using the usual Bohr Sommerfeld rule, while the green dots are obtained by using $\Phi_{Sing}(\epsilon)$. 
The dashed vertical lines represent the interval $|E-E_c| < 2 \hbar \Omega$ of validity of the singular Bohr-Sommerfeld rule.}
\label{bohrsommerfeld}
\end{center}
\nonumber
\end{figure}

The small $x$ analysis gave:
$$
p(x) =  p_+^{\rm Small}(x) + p_-^{\rm Small}(x)
$$
which is valid when $x << 2s_{cl}$. In an intermediate regime, both the small $x$ 
and the WKB approximation are valid at the same time, as can be seen by comparing Eq.~(\ref{psmallx})
and Eq.~(\ref{wkbsmall}). So we can glue these two wave-functions as follows: 
$$
p(x) =  {A_+^{\rm Small}(\epsilon) \over A_+^{\rm WKB}(\epsilon)} p_+^{\rm WKB}(x) +
{A_-^{\rm Small}(\epsilon) \over A_-^{\rm WKB}(\epsilon)} p_-^{\rm WKB}(x)
$$
We can now prolongate the $p_\pm^{\rm WKB}(x)$ functions up to the region $x \simeq \Omega^2-|\xi|$.
Recalling the asymptotic form of the WKB wave function in that region Eq.~(\ref{wkbairy}), we find
\begin{eqnarray*}
p(x) &\simeq&  {A_+^{\rm Small}(\epsilon) \over A_+^{\rm WKB}(\epsilon)} 
{1\over \Omega |\xi|^{1/4}} e^{{2i \over 3 \hbar \kappa} |\xi|^{3/2}
+ i\left( {1\over 2\kappa} - {\epsilon-\kappa \over \Omega^2}\right) |\xi|^{1/2}} \\
&& + {A_-^{\rm Small}(\epsilon) \over A_-^{\rm WKB}(\epsilon)} 
{1\over \Omega |\xi|^{1/4}} e^{-{2i \over 3 \hbar \kappa} |\xi|^{3/2}
- i\left( {1\over 2\kappa} - {\epsilon-\kappa \over \Omega^2}\right) |\xi|^{1/2}}
\end{eqnarray*}
This has to be compatible with the Airy asymptotic formula which embodies the boundary 
conditions Eq.~(\ref{airy}). Hence we find the condition:
\begin{equation}
{ A_+^{\rm Small}(\epsilon) \over  A_-^{\rm Small}(\epsilon) }\;
{A_-^{\rm WKB}(\epsilon) \over A_+^{\rm WKB}(\epsilon) } \; {e^{i{\pi\over 4}} \over e^{-i{\pi\over 4}}}= -1
\label{genbohrsommerfeld}
\end{equation}
This equation determines the energy parameter $\epsilon$ and is the generalized Bohr-Sommerfeld condition.
Given the fact that:
\begin{eqnarray*}
{A^{\rm Small}_+(\epsilon)\over A^{\rm Small}_-(\epsilon)} &=& e^{i(\nu-{\pi\over 2})}
{ \Gamma \left( {1\over 2} -i {\epsilon - \kappa \over 2 \Omega} \right) \over 
\Gamma \left( {1\over 2} +i {\epsilon - \kappa \over 2 \Omega} \right) } 
e^{i {\epsilon -\kappa \over \Omega} \log \left({ 2 \Omega \over \sqrt{2s_{cl}}}\right)} \\
{A^{\rm WKB}_+(\epsilon)\over A^{\rm WKB}_-(\epsilon)} &=& e^{-2 i \left[ { 2\kappa \Omega \over \hbar }
+(4s+1)\nu+{\epsilon-\kappa \over \Omega} \log \left( {4 \Omega^2 \over \hbar} \right) \right]}
\end{eqnarray*}
Eq.(\ref{genbohrsommerfeld}) becomes:
\begin{equation}
{ \Gamma \left( {1\over 2} -i {\epsilon - \kappa \over 2 \Omega} \right) \over 
\Gamma \left( {1\over 2} +i {\epsilon - \kappa \over 2 \Omega} \right) } \;  
e^{i\left[{ 2\kappa \Omega \over \hbar}+2(2s+1)\nu+
{\epsilon-\kappa \over \Omega} \log \left({8\Omega^3 \over \hbar \sqrt{2s_{cl}}}\right)\right] } =-1
 \end{equation}
Taking the logarithm, we find the quantization condition:
\begin{equation}
 \Phi_{Sing}(\epsilon_n)= 2\pi \left( n +{1\over 2} \right), \quad n\in Z, \quad E_n = 2\kappa s_{cl} + \hbar \epsilon_n
 \label{SBS}
 \end{equation}
where: 
$$
\Phi_{Sing}(\epsilon)= -i \log  { \Gamma \left( {1\over 2} -i {\epsilon - \kappa \over 2 \Omega} \right) \over 
\Gamma \left( {1\over 2} +i {\epsilon - \kappa \over 2 \Omega} \right) }  + 
{ 2\kappa \Omega \over \hbar}+2(2s+1)\nu+
{\epsilon-\kappa \over \Omega} \log \left({8\Omega^3 \over \hbar \sqrt{2s_{cl}}}\right) 
$$
To test this condition, we can compute $\delta_n= \Phi(\epsilon_n^{\rm exact})- 2\pi \left( n +{1\over 2} \right)$ 
for the exact values of the energies. In Figure~\ref{bohrsommerfeld}, we plot $\delta_n$ as a function of $n$, 
using for $\Phi(\epsilon)$ both the usual Bohr-Sommerfeld function:
$$
\Phi_{Reg}(\epsilon) = {1\over \hbar }\int_{C(E)} \alpha + \int_{C(E)} \gamma
$$
and the function $\Phi_{Sing}(\epsilon)$. Near the singularity, the singular Bohr-Sommerfeld condition is much 
more accurate.

\begin{figure}[hbtp]
\begin{center}
\includegraphics[height= 5cm,width=15cm]{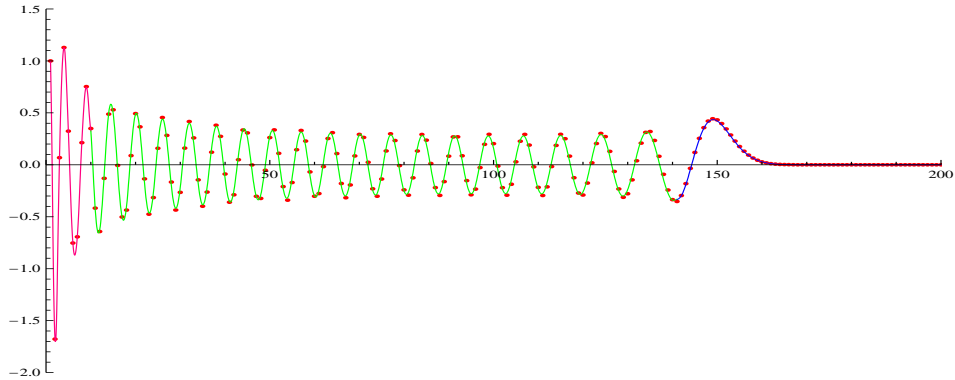} 
\caption{The components of an eigenvector close to the critical level. The dots are the exact result. The red curve is the result of the small $x$ analysis. The green curve is the WKB result and the blue curve is the Airy result.}
\label{superposition}
\end{center}
\nonumber
\end{figure}

In Figure~\ref{superposition} a typical example of the components of the eigenvectors  is shown, comparing with the exact result obtained by direct diagonalization of the Jacobi matrix Eq.~(\ref{EigEq}) and the various results corresponding to the different 
approximations: small $x$, WKB, and Airy. The agreement is very good.

\begin{figure}[hbtp]
\begin{center}
\includegraphics[height= 5cm,width=15cm]{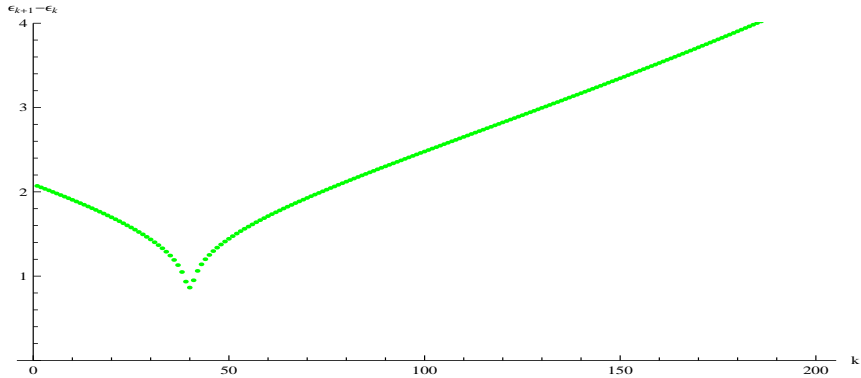} 
\caption{The energy level spacing $\epsilon_{k+1}-\epsilon_{k}$
as a function the the state label $k$. This plot shows clearly the accumulation of energy levels in the vicinity
of the critical point.}
\label{levelspacing}
\end{center}
\nonumber
\end{figure}

From Eq.~(\ref{SBS}) we can compute the level spacing between two successive energy levels. It is given by:
\begin{equation}
\epsilon_{n+1}-\epsilon_n = {2\pi\over \Phi_{Sing}'(\epsilon_n)}= {2\pi \Omega \over  
\log{\left(8\Omega^3\over \hbar\sqrt{2s_{cl}}\right)} - \Psi'({\epsilon-\kappa \over 2\Omega})}
\label{depsilonn}
\end{equation}
where we defined
$$
\Psi'(x) ={i \over 2} {d\over dx} \log {\Gamma \left( {1\over 2} - ix \right) \over \Gamma \left( {1\over 2} + ix \right)}
$$
This function has a sharp minimum located at $x=0$ where it takes the value $-(\gamma + 2 \log 2)$ where $\gamma$ is Euler's constant, 
see Fig.~\ref{levelspacing}.
Hence to leading order, the smallest energy spacing is:
$$
\Delta \epsilon \simeq {2\pi\Omega \over |\log \hbar |}
$$
A detailed analysis of the level spacing in the vicinity of the critical level has been given recently~\cite{Keeling09},
where it has been applied to the description of the long time dynamics of the system.

\section{ Evolution of the oscillator energy}
\label{evolution}
\begin{figure}[hbtp]
\begin{center}
\includegraphics[height= 5cm,width=15cm]{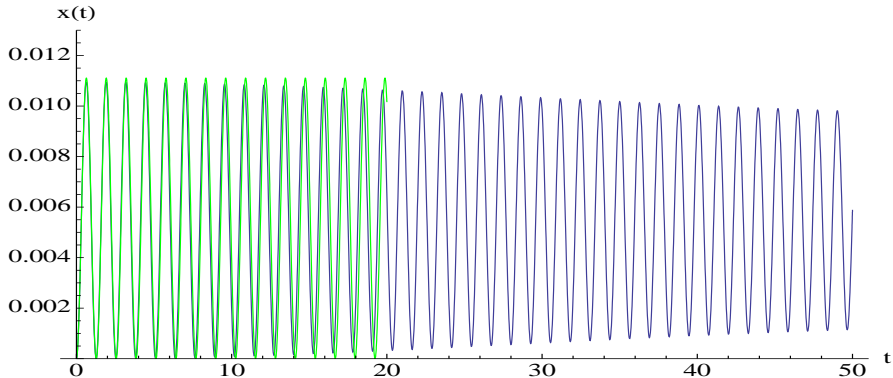} 
\caption{The excitation number of the oscillator as function of time in the stable case. 
The green curve is the small time result Eq.~(\ref{xbarstable}).
($\hbar^{-1}=s=30$,  $\kappa= -2\sqrt{2s_{cl}}$,
$s_{cl}=1$). Notice that  the vertical amplitude is about one hundred times smaller than in the unstable case (compare with Fig.~\ref{molecules}).}
\label{nomolecules}
\end{center}
\nonumber
\end{figure}

Once the eigenvectors and  eigenvalues are known, we can compute the time evolution of the oscillator energy:
$$
\bar{x}(t) = \langle s \vert e^{i{t H\over \hbar}} b^\dag b e^{-i{ tH\over \hbar}} \vert s \rangle
$$
Since the observable $b^\dag b$ commutes with $H_1$, its  matrix elements between eigenstates of $H_1$ 
with different eigenvalues vanish. Because $\vert s \rangle$ is an eigenvector of $H_1$ with eigenvalue $\hbar s$, 
we can therefore restrict ourselves to this eigenspace of $H_1$.
The time evolution can be computed numerically by first decomposing the initial state on the
eigenvector basis:
\begin{equation}
\vert s \rangle = \sum_{n }  c_n \vert \Psi(E_n)\rangle
\label{defcn}
\end{equation}
so that the wave-function at time $t$ is:
$$ 
\vert \Psi(t) \rangle = \sum_{n }  c_ne^{-iE_n t/\hbar} \vert \Psi(E_n)\rangle
$$
In the stable regime $\kappa^2 > 2s_{cl} $, we get quite regular oscillations with
a rather small amplitude as shown in Figure~\ref{nomolecules}.

\begin{figure}[hbtp]
\begin{center}
\includegraphics[height= 5cm, width=15cm]{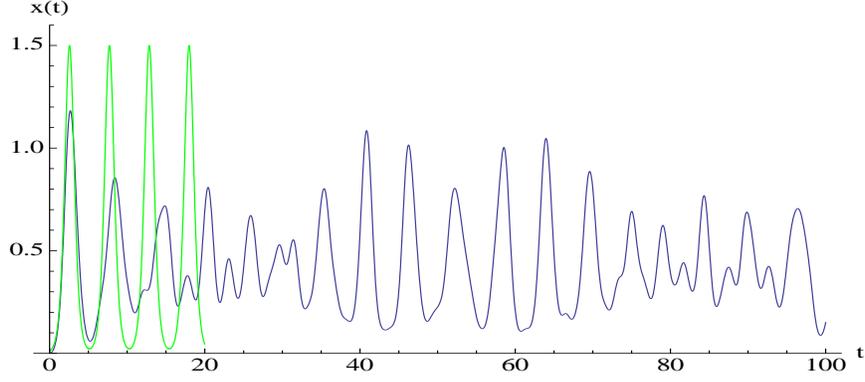} 
\caption{The excitation number of the oscillator as function of time in the unstable case. ($\hbar^{-1}=s=30$,  $\kappa= -0.5\sqrt{2s_{cl}}$,
$s_{cl}=1$). It clearly shows an aperidodic succession of pulses. The green curve is the semiclassical result Eq.~(\ref{xbarinstable}). }
\label{molecules}
\end{center}
\nonumber
\end{figure}

By contrast, in the unstable regime $\kappa^2 < 2s_{cl}$, it is energetically favorable to excite the oscillator, 
and we get a succession of well separated pulses shown in Figure~\ref{molecules}.
Note that the temporal succession of these pulses displays a rather well defined periodicity, but
the fluctuations from one pulse to another are relatively large.  
The remaining part of this paper is an attempt to understand the main features of this evolution.

\begin{figure}[hbtp]
\begin{center}
\includegraphics[height= 5cm,width=15cm]{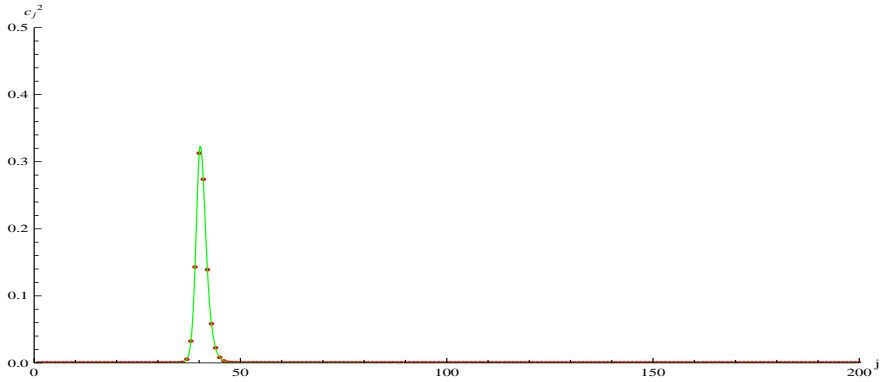} 
\caption{The coefficients of the expansion of the initial state $|s\rangle$ on the eigenvectors $\vert \Psi(E_n)\rangle$. 
The red dots are the exact values. We see that only eigenvectors close to the critical level contribute significantly.
The curve is given by Eq.~(\ref{ce}), which is deduced from the knowledge of the
short time evolution. The agreement between this approximate result and the exact one is remarkable, even in the
immediate vicinity of the critical level.}
\label{figeigencoeff}
\end{center}
\nonumber
\end{figure}

On Figure~\ref{figeigencoeff} we show the coefficients $|c_n|$. Only those eigenstates whose energy 
is close to the critical classical energy  $E_c$ contribute significantly. A good estimate of the
energy width of the initial state $|s\rangle$ is given by 
$\Delta E=\left( \langle s|H^{2}|s \rangle - \langle s|H|s \rangle^{2}\right)^{1/2}$.
From Eq.~(\ref{EigEq}), we find:
\begin{equation} 
\Delta E= \hbar \sqrt{2s_{cl}}=\hbar \Omega /\sin\nu
\label{deltaE}
\end{equation} 
Comparing with the criterion $|\Delta E| \gg 2\hbar \Omega$ for the validity of the stationary phase 
approximation, we see that most of the eigenstates which have a significant weight in the spectral
decomposition of the initial state actually belong to the singular Bohr-Sommerfeld regime.

\begin{figure}[hbtp]
\begin{center}
\includegraphics[width=13cm]{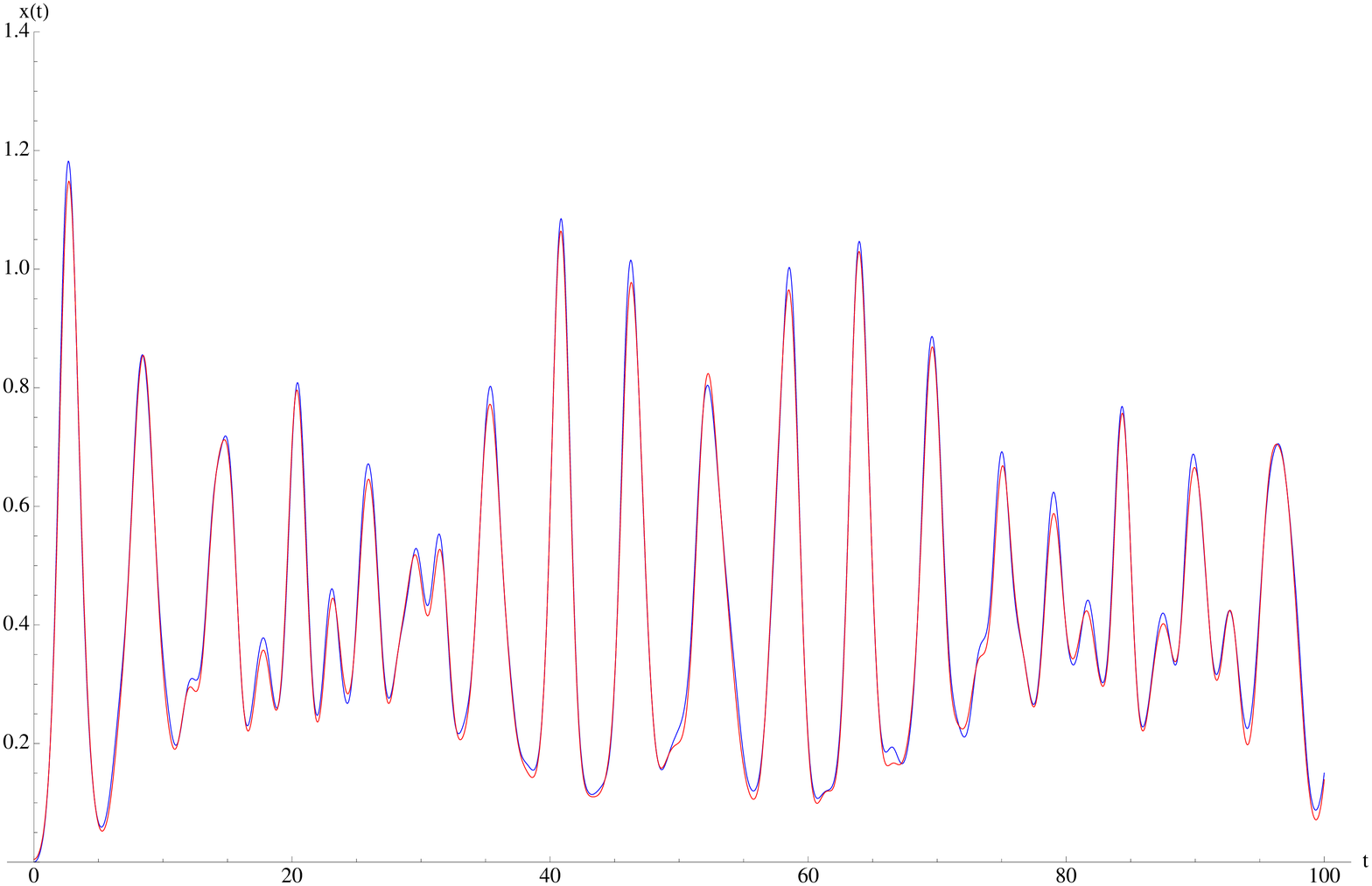} 
\caption{The excitation number of the oscillator as function of time for $s=30$.
The blue curve is the exact result, and the red curve is obtained by keeping only 6 energy eigenstates.}
\label{nt6}
\end{center}
\nonumber
\end{figure}

An important consequence of this observation is that we can compute  $\bar{x}(t)$ 
by considering only the few relevant states. 
The result is shown in Figure~\ref{nt6} for a spin $s=30$. 
We retained only six states and superposed the result to the exact curve obtained by keeping the 61 states. 
We can hardly see any difference between the two curves.

We can compute the coefficients $|c_n|$ of the decomposition of the initial state $|s\rangle$. From Eq.~(\ref{defcn}), we have:
\begin{equation}
\langle s | e^{-i{Ht\over \hbar}} | s \rangle = \sum_n e^{-i{E_n t \over \hbar}} |c_n|^2 \simeq 
\int_{-\infty}^{\infty} e^{-i{Et\over \hbar}} |c(E)|^2 \rho(E) dE
\label{defce}
\end{equation}
where we have approximated the sum over discrete energy levels by an integral. Since we already know that the integral is concentrated around the critical energy $E_c$, the density $\rho(E)$ can be computed from the singular Bohr-Sommerfeld phase eq.(\ref{SBS}):
$$
\rho(E) = {dn\over dE} = {1\over 2\pi } {d \over dE} \Phi_{Sing}(E)= {1\over 2\pi \hbar} {d \over d\epsilon} \Phi_{Sing}(\epsilon)
$$
On the other hand, since
$$
|s\rangle = {e_0\over ||e_0||}
$$
where $e_0$ is the state Eq.~(\ref{defen}) with the oscillator in its ground state, we get: 
$$
\langle s | e^{-i{Ht\over \hbar}} | s \rangle = {1\over 2\pi} \int_0^{2\pi} \Psi(\theta,t) d\theta =p_0(t)
$$
where $\Psi(\theta,t)$ is the solution of the Schr\"odinger equation with initial condition
$\Psi(\theta,t)\vert_{t=0}=1$. Inverting the Fourier transform in Eq.~(\ref{defce}), we arrive at:
$$
|c(E)|^2 \rho(E) = {1\over 2\pi \hbar} \int_{-\infty}^\infty e^{i{E t\over \hbar}} p_0(t) dt
$$
Taking for $p_0(t)$ the small time expression Eq.~(\ref{pnt}), which is valid for times $t< t_0 \simeq {|\log \hbar |\over 2\Omega}$,
we expect to find an expression valid in the energy interval $|E-E_c| > 2 \Omega {\hbar \over |\log \hbar |}$ i.e. for energies 
far enough from the critical energy. However because of the factor $1/|\log \hbar |$ this interval covers most of the range Eq.~(\ref{deltaE}) 
where we expect $c(E)$ to be substantially non zero. Notice also that $\hbar\over |\log \hbar | $ is  the order of magnitude of level spacing  
in the critical region, so that the interval where the use of the small time wave function is not legitimate contains at most a few levels. 
Explicitly,  computing the Fourier transform of Eq.~(\ref{pnt}), we find:
\begin{equation}
|c(E)|^2 \rho(E) ={1\over \hbar \sqrt{2 s_{cl}}} {e^{-{ \nu  \alpha(E)\over \hbar \Omega}} \over 1+ e^{-{\pi \alpha(E) \over \hbar \Omega}}}
\label{ce}
\end{equation}
where: 
$$
\alpha(E) = E-E_c - \hbar \kappa
$$
The coefficients $|c(E)|^2$ computed with this formula are shown in Fig.~\ref{figeigencoeff}. The agreement with the exact coefficients is excellent.

\bigskip

For longer time scales, we may also infer that the quantum dynamics exhibits the Bohr frequencies that can be
deduced from the singular Bohr-Sommerfeld quantization rule~(\ref{SBS}). As shown before, the typical
spacing between these energy levels is given by  $\Delta \epsilon \simeq {2\pi\Omega \over |\log \hbar |}$.
This corresponds to a fundamental period $\Delta t \simeq |\log \hbar|/\Omega$. The fact that the energy
levels are not equally spaced, as shown on Figure~\ref{levelspacing} is responsible for the aperiodic
behavior on time scales larger than $\Delta t$. Systematic procedures to analyze the long time behavior,
starting from a nearly equidistant spectrum have been developed~\cite{Leichtle96}, but we have not yet
attempted to apply them to the present problem. Note that the quantum dynamics in the vicinity of
classically unstable equilibria has received a lot of attention in recent years~\cite{Micheli03,Boukobza09}. 

\bigskip

Instead, motivated by the analysis of the previous section,  we now present a discussion of the time evolution in the semi-classical regime. As for the stationary levels, we show that we have to treat separately the initial evolution, when $x$ is still
small and the system is close to the unstable point, and the motion at later times, for which the
usual WBK approach is quite reliable. As we demonstrate, this allows us to predict the appearance of pulses,
together with their main period $\Delta t$. This generalizes the early study by  
Bonifacio and Preparata~\cite{Bonifacio}, who focussed on the $\kappa=0$ case.

\subsection{Small time analysis}

Another expression for the mean oscillator energy is:
$$
\bar{x}(t) =  \hbar \sum_{n=0}^{2s} n |p_n(t)|^2
$$
where $p_n(t)$ is the solution of Eq.~(\ref{schroerduite}) with boundary condition:
$$
p_n(t)\vert_{t=0} = \delta_{n,0}
$$
For small time, $p_n(t)$ will be significantly different from zero only for small $n$. 
So in Eq.~(\ref{schroerduite}) we may assume $n << 2s$. The equation becomes:
\begin{equation}
i\hbar {\partial p_n \over \partial t}= \hbar \sqrt{2s \hbar} (n+1) p_{n+1}  + \hbar  \sqrt{2s\hbar } \;n p_{n-1} 
 + 2\hbar \kappa (s-n)  p_n 
 \label{schroerduite2}
\end{equation}
As for the stationary case, this equation is linear in $n$ and can be solved by Laplace-Fourier transform. 
We define:
\begin{equation}
\Psi(\theta, t) = \sum_{n=0}^\infty p_n(t) e^{in \theta}, \quad \Psi(\theta, t)\vert_{t=0} = 1
\label{Psidef2}
\end{equation}
The time evolution is given by: 
\begin{equation}
\partial_t \Psi =-( \sqrt{2s_{cl}}(e^{i\theta} + e^{-i\theta} ) -2\kappa) \partial_{\theta} \Psi
-i (2 \kappa s + \sqrt{2s_{cl} }e^{i\theta} ) \Psi
\label{schroetheta2}
\end{equation}
This equation can be solved by the method of characteristics. We introduce
the function $y(\theta)$ defined by: 
$$
{d y \over d\theta}= {1\over  \sqrt{2s_{cl}}(e^{i\theta} + e^{-i\theta} ) -2\kappa}
$$
Explicitely:
$$
y(\theta) = {1\over 2 \Omega} \log { e^{i\theta} + e^{i \nu} \over  e^{i\theta} + e^{-i \nu}}, \quad {\rm or~else} \quad e^{i\theta} = {1\over \sqrt{2s_{cl}}} \left(\kappa + i\Omega {\cosh \Omega y \over \sinh \Omega y}\right)
$$
The time dependent Schr\"odinger equation becomes:
$$
\partial_t \Psi + \partial_y \Psi =  - i(2 s+1)\kappa \Psi + \Omega  {\cosh \Omega y \over \sinh \Omega y} \Psi 
$$
whose solution reads: 
$$
\Psi(y,t) = e^{-i\kappa(2s+1)t } {\sinh \Omega y \over \sinh \Omega(y-t)} \Psi(y-t,0)
$$
Imposing the initial condition Eq.~(\ref{Psidef2}) yields $\Psi(y-t,0)=1$ and then:
$$
\Psi(y,t) = e^{-i\kappa(2s+1)t } {\sinh \Omega y \over \sinh \Omega(y-t)}
$$
Returning to the variable $\theta$, we get:
\begin{equation}
\Psi(\theta,t) = e^{-i\kappa(2s+1)t } \; {e^{i\nu}-e^{-i\nu} \over e^{i\nu-\Omega t} -e^{-i\nu+\Omega t}}\;
 {1\over 1-{e^{\Omega t} - e^{-\Omega t} \over e^{i\nu - \Omega t} - e^{-i\nu + \Omega t}} \; e^{i\theta}}
\label{solution_temps_courts}
\end{equation}

As for the stationary case, we note that this wave-function has exactly the form dictated by
semi-classical analysis, where the full symbol associated to
Eqs.~(\ref{schroerduite2}),(\ref{schroetheta2}) is 
$h(x,\theta)=2\sqrt{2s_{cl}}(x+\hbar/2)(\cos\theta+\cos\nu)$, after we have subtracted the energy
$E_c$ of the unstable point.
The generalization of Eq.~(\ref{fonctiondephase}) to the time dependent case is: 
\begin{equation}
h^{0}(\partial_{\theta}S_{0}(\theta,t),\theta) + \partial_{t}S_{0}(\theta,t) = 0  
\end{equation}
But since $\Psi$ is independent of $\theta$ at $t=0$, $\partial_{\theta}S_{0}(\theta,t=0)=0$, which
means classically that the trajectory begins at $x=0$. But because of the form of $h^{0}$, we
get $h^{0}(\partial_{\theta}S_{0}(\theta,t=0),\theta)=0$ for any $\theta$, which implies that
$\partial_{t}S_{0}(\theta,t=0) = 0$. This shows that $S_{0}(\theta,t) = 0$ identically, in agreement
with the fact that a classical trajectory starting at $x=0$ stays there for ever!
So the time evolution manifested in Eq.~(\ref{solution_temps_courts}) is of purely quantum nature,
and it is encoded in the evolution of the amplitude $a(\theta,t)$ and the subleading phase $S_{1}(\theta,t)$.
Writing $\Psi=a(\theta,t)\exp(iS_{1}(\theta,t))$, the next order in $\hbar$ gives:
\begin{equation}
\partial_{x}h^{0}(0,\theta)\partial_{\theta}\Psi+\partial_{t}\Psi+\frac{1}{2}\frac{d}{d\theta}
\left(\partial_{x}h^{0}(0,\theta)\right)_{|t}\Psi+ih^{1}(0,\theta)\Psi=0
\end{equation}
A simple check shows this is exactly the same equation as (\ref{schroetheta2}) without the
term $-i2\kappa s\Psi$, due to the subtraction of the energy $E_c$.
Expanding in $e^{i\theta}$ we find:
\begin{equation}
p_n(t) =e^{-i\kappa(2s+1)t } \; {e^{i\nu}-e^{-i\nu} \over e^{i\nu-\Omega t} -e^{-i\nu+\Omega t}}
\; \left[ {e^{\Omega t} - e^{-\Omega t} \over e^{i\nu - \Omega t} - e^{-i\nu + \Omega t}}\right]^n
\label{pnt}
\end{equation}
It is instructive to consider the large $n$ limit. Then $p_{n}(t)$
is proportional to $\exp[-\frac{i}{\hbar}(\kappa (2s_{cl}+\hbar)t+(\pi-\nu)x)]$. This is the 
dominant phase factor for a WKB state of energy $E=E_c$ which is concentrated on the
outgoing branch of the critical classical trajectory. Note that there also appears a quantum
correction to the energy, corresponding to a finite $\epsilon=\kappa$. On the physical side, this
is quite remarkable, because we have just seen that the short time evolution is driven by
purely quantum fluctuations, formally described by the subprincipal symbol. Nevertheless, the subsequent
evolution is quite close to the classical critical trajectory, because the energy distribution of
the initial state is quite narrow, as we have discussed.

Further information is obtained by looking at the probability distribution of the number of emitted quantas:
$$
| p_n(t)|^2 = {\Omega^2 \over \Omega^2 + 2 s_{cl} \sinh^2 \Omega t } 
\left[ {2 s_{cl} \sinh^2\Omega t \over \Omega^2 + 2 s_{cl} \sinh^2 \Omega t }  \right]^n
$$
This is of the form:
$$
| p_n(t)|^2 \simeq e^{-\beta(t) n}, \quad \beta(t)= \log \left( 1 + {\Omega^2 \over 2s_{cl} \sinh^2 \Omega t} \right)
$$
Hence we have a {\em thermal} distribution with a time-dependent effective temperature.
Such behavior is due to the strong entanglement between the spin and the oscillator.
In fact we somehow expect that quantum effects, such as entanglement, are likely to be
magnified in situations where there is an important qualitative difference between the classical
and the quantum evolutions.

It is now simple to compute the mean number of molecules produced in this small time regime. We find:
\begin{equation}
\bar{x}(t) =2  \hbar s_{cl} {\sinh^2\Omega t \over \Omega^2}, \quad \Omega = \sqrt{2s_{cl}-\kappa^2}
\label{xsmalltime}
\end{equation}
The small time approximation is valid as long as the number of molecules is small and will break after a time  $t_0$ such that $\bar{x}(t_0)\simeq 1$. To leading order in $\hbar$, this time scale is:
$$
t_0 \simeq -{1\over 2 \Omega} \log \hbar
$$
Note that in the {\em stable} case, we change  $\Omega$ into $i\Omega$ and then: 
\begin{equation}
\bar{x}(t) =2  \hbar s_{cl} {\sin^2\Omega t \over \Omega^2}, \quad \Omega = \sqrt{\kappa^2 - 2 s_{cl}}
\label{xbarstable}
\end{equation}
In that case we never leave the small $n$ regime and this approximation remains valid even for large time as can be seen on Fig.~\ref{nomolecules}.

\subsection{Periodic Soliton Pulses}

When time increases, the approximation $n << 1$ is not valid anymore and we must take into account the exact
quantum Hamiltonian. In the regime $\hbar << x = n\hbar << 2s_{cl}$, we can still perform a WKB approximation. 
The previous discussion has shown that the semi classical wave-function is concentrated on the classical
orbit of $h^{0}$ with energy $E=E_c+\hbar\kappa$, that is $\epsilon = \kappa$. 
We set:
$$
p(x,t) = e^{- {i\over \hbar} \kappa (2s_{cl}+\hbar) \; t}  e^{-{i\over \hbar} (W_{0}(x)+\hbar W_{1}(x)} b(x,t)
$$
where $W_{0}(x)$ is given by Eq.~(\ref{soluWKB}), and $W_{1}(x)$ by Eq.~(\ref{expressionW1}) in which 
$\epsilon$ is replaced by it actual value $\kappa$. The only source of time dependence, at this level
of approximation, arises from the transport equation of the amplitude. Let us define the local velocity
$v(x)$ on the classical trajectory by:
$$
v(x)=-\partial_{\theta} h^{0}(x,W'_{0}(x)) = \pm 2x \sqrt{\Omega^{2}-x}
$$
From this field, we obtain the Hamiltonian flow on the classical trajectory
by solving the differential equation $dx/dt=v(x)$. Let us denote by
$x(x_{0},t)$ the solution of this equation which starts from $x_{0}$ at $t=0$.
Likewise, we may reverse the flow and define $x_{0}(x,t)$.
It is convenient to introduce the function $u(x)$ such that $du/dx=-1/v(x)$.
The solution of the flow is then:
\begin{equation}
u(x_{0})=u(x)+t
\end{equation}
For our problem, we have:
$$
u(x) = - \int^x {1\over 2x \sqrt{\Omega^2 -x}} dx, \quad {\rm or~else} \quad x= {\Omega^2 \over \cosh^2 \Omega u}
$$

The transport equation simply states the conservation of the local density $b^{2}(x,t)$:
$$
\partial_{t}(b^{2}(x,t))+\partial_{x}(b^{2}(x,t)v(x))=0
$$
The evolution of this one dimensional conserved fluid is characterized by
the invariance of $b^{2}v$ along the flow:
\begin{equation} 
b^{2}(x,t)v(x)=b^{2}(x_{0}(x,t),0)v(x_{0}(x,t)) \equiv B_{0}^{2}(u(x)+t)
\end{equation}
which simply results from the conservation law and the fact that the velocity field 
does not depend on time.   

\begin{figure}[hbtp]
\begin{center}
\includegraphics[height= 5cm, width=10cm]{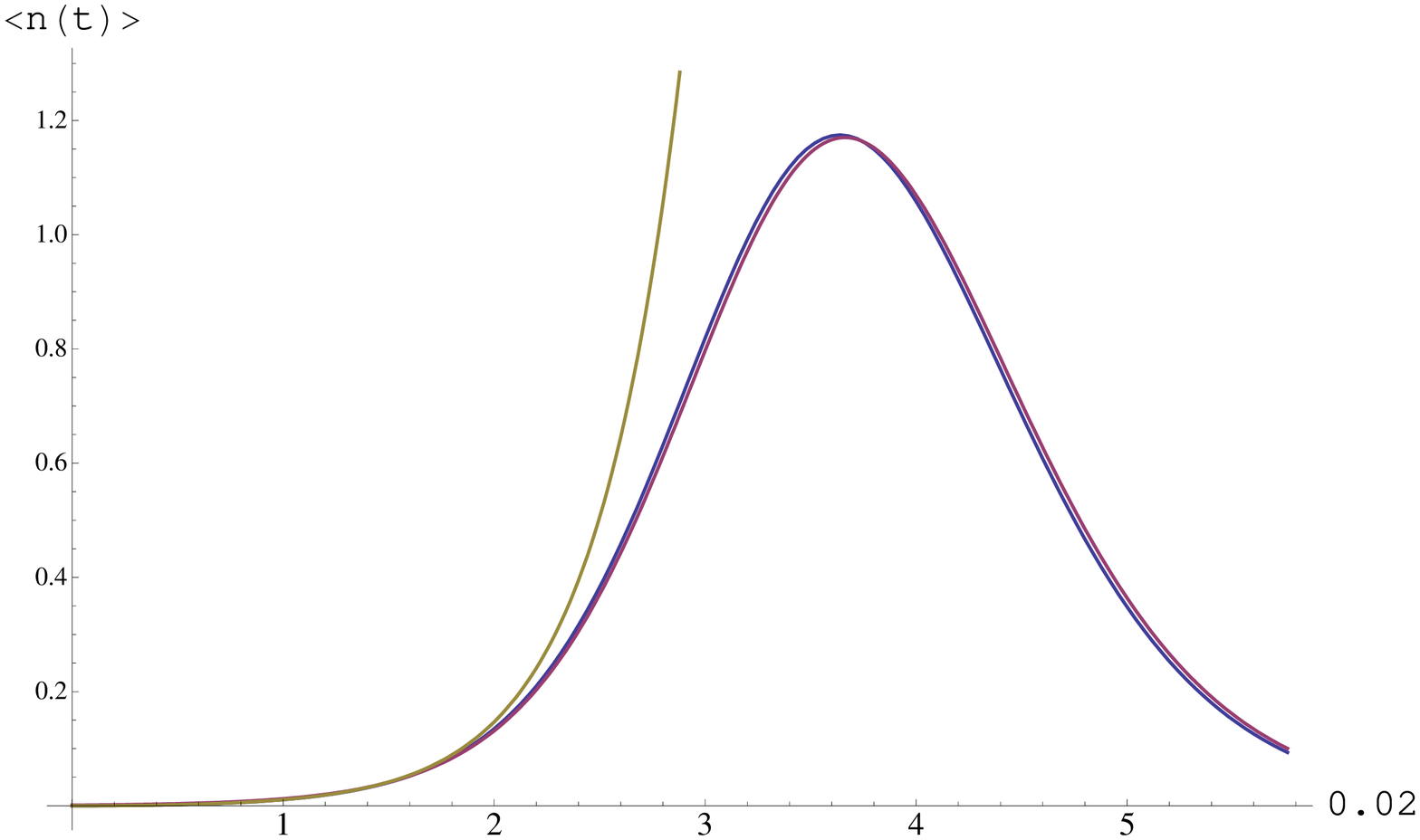} 
\caption{Time evolution of the mean oscillator energy
 $\bar x(t)$ for $\kappa=1/\sqrt{2}$, $s_{cl}=1$ and
$\hbar=1/300$. The blue line corresponds to the exact solution, the
red one to the semiclassical solution and the
yellow line to the small time solution.}
\label{first-peack}
\end{center}
\end{figure}

Then we have:
\begin{equation}\label{semicl-evol-int}
\bar{x}(t) = \int dx  {x\over [ x^2 (\Omega^2 -x)]^{1/2} } B_0(u(x)+t)^2 =
\int du {\Omega^2 \over \cosh^2 \Omega u} B_0(u+t)^2 
\end{equation}
If we assume that $B_0(u)^2$ is peaked around $t_0$, say $B_0(u)^2 = \delta(u-t_0)$, we find:
\begin{equation}
\bar{x}(t) =  {\Omega^2 \over \cosh^2 \Omega (t-t_0)}
\label{xbarinstable}
\end{equation}
which is just the classical expression. There is a simple way to
evaluate the time $t_0$. In fact there must be a regime where the
small time formula and this semiclassical formula should agree. i.e. 
\begin{equation}
\hbar {2s_{cl} \over \Omega^2} \sinh^2\Omega t \simeq {\Omega^2  \over
  \cosh^2 \Omega (t-t_0)}, \quad  
{\rm or} \quad \hbar {2s_{cl} \over 4 \Omega^2} e^{2 \Omega t } \simeq 4
\Omega^2  e^{2 \Omega (t-t_0) }  
\end{equation}
and this gives the time $t_0$:
\begin{equation}\label{t0}
t_0 = -{1\over 2 \Omega} \log{ 2 s_{cl} \over 16 \Omega^4} \hbar
\simeq -{1\over 2 \Omega} \log \hbar + O(\hbar^0)
\end{equation}
This reproduces well the position of the first peak. Its height is also well
reproduced if instead of the crude approximation $B_0(u)^2 =
\delta(u-t_0)$, we glue the wave functions given by the small time solution and the WKB solution at a gluing time $t_g$:
$$
B_0(u + t_g)^2 \simeq K \frac{\sinh(\Omega u)}{\cosh(\Omega u)^3}
\exp\left(-\frac{\Omega^2}{\bar x(t_g) \cosh(\Omega u)^2 }\right)
$$
where  the normalization constant $K$ is given by the condition:
$$
\int_0^\infty B_0(u)^2 du =1
$$
and $\bar x(t_g)$ is the value of $\bar x$  given by Eq.~(\ref{xsmalltime}) taken at time $ t=t_g$ (see Fig.~\ref{first-peack}).
The height of the first peak can be estimated from
Eq.~(\ref{semicl-evol-int}). The maximum is obtained when we have the
largest overlap between $1/\cosh^2{\Omega u}$ and $
B_0(u+t)^2$. This is happens with a very good approximation, for 
$t=t_0$ such that the maximum of $B_0(u+t_0)^2$ is at zero:
$$
\sinh^2{\Omega (t_0-t_g)} = -\frac{1}{4}\left(1-\frac{2\Omega }{\bar x(t_g)}
\right)+ \sqrt{\frac{1}{16}\left(1-\frac{2\Omega}{\bar x(t_g)}
  \right)^2+ \frac{1}{2}}  
$$

\begin{figure}[hbtp]
\begin{center}
  \includegraphics[height= 5cm, width=10cm]{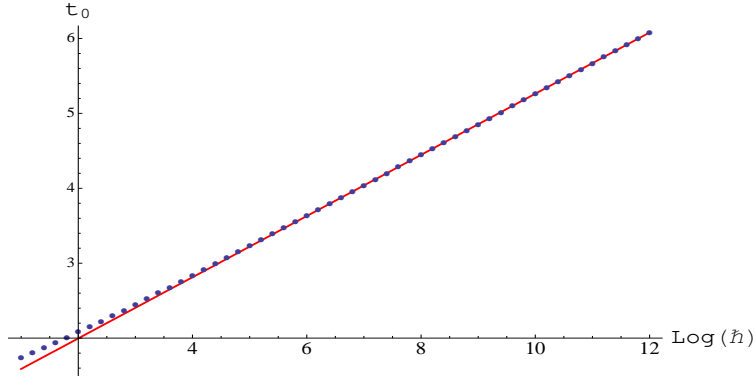} 
\caption{Time $t_0$ of the first pulse as a function of $\hbar$.
The blue points are obtained by looking at the value of $t_0$ at the
plateau. The red line corresponds to the Eq.~(\ref{t0}). $\kappa=-1/\sqrt{2}$, $s_{cl}=1$.
}
\label{t0fig}
\end{center}
\nonumber
\end{figure}

The previous equation allows us also to discuss the freedom in the
choice of the gluing point $t_g$. Indeed we expect that $t_0$ does not
depend on $t_g$ as long as $t_g$ is restricted to a temporal domain where the
short time solution and the semiclassical one overlap. If we plot $t_0$ as a
function of $t_g$, we see that the curve we obtain develops a plateau,
which means that the value of $t_0$ we obtain chosing $t_g$ inside this plateau is independent of $t_g$. 
\begin{center}
\begin{figure}
\includegraphics[height= 5cm, width=10cm]{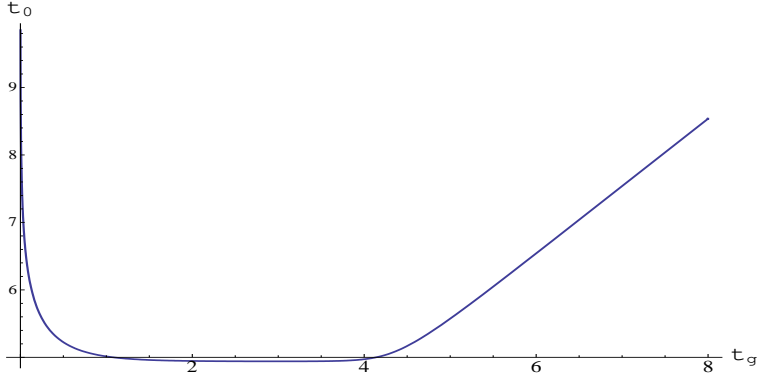} 
\caption{Time $t_0$ of the first pulse as a function of the gluing time $t_g$ between
the short time solution and the WKB solution. The existence of a plateau shows that gluing
can be done in an unambiguous way.  $\kappa =-1/\sqrt{2}$, $s_{cl}=1$ and
$\hbar=10^{-4}$. We can recognize a plateau between $t_g=1$ and $t_g=4$.}
\end{figure}
\end{center}
Moreover, if we plot the value of $t_0$ at the plateau, as a function of $\log(\hbar)$, we
recover with a very good accuracy the rough estimation of
Eq.~(\ref{t0}), see Fig.~\ref{t0fig}.

If we return to the $delta$-function approximation, it is clear that the situation reproduces itself after a time interval $2 t_0$. 
Hence we can write:
$$
\bar{x}(t) \simeq \sum_{k=-\infty}^{\infty} {\Omega^2 \over \cosh^2 \Omega(t -(2k+1)t_0)} \simeq 
\Omega^2 {\rm cn}^2[\Omega(t-t_0) \vert k^2]
$$
where ${\rm cn}$ is a Jacobi elliptic function of modulus $k^2$ given approximately by: 
$$
k^2 \simeq 1-{2\hbar s_{cl} \over \Omega^4}
$$
When $\kappa =0$, this is Bonifacio and Preparata's result~\cite{Bonifacio}.
Clearly, it remains a challenge to account for the lack of exact periodicity in the pulse sequence, using the time-dependent approach.  In this perspective, returning to the integrable structure may be interesting in order to obtain an effective model around the critical point  in terms of particle-hole excitations of Bethe pseudo-particles.

\section{Conclusion}

Let us emphasize the main results of this work. First, the Jaynes-Cummings model exhibits, in  a large region of its
parameter space, an unstable fixed point which corresponds to the focus-focus singularity of an integrable system
with two degrees of freedom. As a result, there is a classical monodromy when one considers a loop which encircles the
critical value in the $(H_{0},H_{1})$ plane. At the quantum level, this phenomenon is manifested by a dislocation in the
joint spectrum of $H_{0}$ and $H_{1}$. We have then analyzed the eigen-subspace of $H_{1}$ which corresponds to the
critical point. The associated reduced phase-space is a sphere. We have shown how to perform a semi-classical analysis
using the convenient but singular coordinates $(x,\theta)$ on this sphere.
The main result here is that when the classical orbit crosses either the north or the south pole, where the longitude
$\theta$ is not well defined, the action integral associated to the subprincipal symbol jumps by $\pm \pi$, and this
is compensated by a simultaneous jump in the Maslov index. Most of the spectrum is well described by usual Bohr-Sommerfeld
quantization rules, at the exception of typically $|\log \hbar|$ eigenstates in the vicinity of the critical energy, for which
special Bohr-Sommerfeld rules have been obtained. Remarkably, the classical unstable equilibrium state, where the spin component
$s^{z}$ is maximal and the oscillator is in its ground-state, has most of its weight on the subspace spanned by these singular
semi-classical states. This fact explains rather well the three time scales observed in the evolution of the mean energy of the
oscillator. At short time, this energy grows exponentially, reflecting the classical instability of the initial condition.
At intermediate times, the energy of the oscillator exhibits a periodic sequence of pulses, which are well described by the
classical motion along the pinched torus containing the unstable point. Finally, the delicate pattern of energy levels, which are
not exactly equidistant, governs the aperiodic behavior observed for longer time scales.

This work leaves many unsolved questions. One of them is to develop a detailed analytical description of the
long time behavior. This should a priori be possible because, as we have seen, the initial state is a linear
superposition of only a small number of energy eigenstates, for which the singular WKB analysis developed here
provides an accurate modeling. Another interesting direction is the extension to several spins, which is
physically relevant to the dynamics of cold atom systems after a fast sweep through a Feshbach resonance.
It would be interesting to discuss if the notion of monodromy can be generalized with more than two degrees of freedom.
Finally, it remains to see whether the qualitative features of the time evolution starting from the unstable state
remain valid for an arbitrary number of spins, as may be conjectured from the structure of the quadratic normal form
in the vicinity of the singularity.

\section{Acknowledgements}

One of us (B.D.) would like to thank Yshai Avishai for many stimulating discussions on out of equilibrium dynamics
of cold atom systems and for an initial collaboration on this project. We have also benefited a lot from many
extended discussions with Thierry Paul on various aspects of semi-classical quantization. Finally, we wish to thank
Yves Colin de Verdi\`ere and Laurent Charles for sharing their knowledge with us.
B.D.was supported in part by the ANR Program QPPRJCCQ, ANR-06BLAN-0218-01, 
O.B. was partially supported by the European Network  ENIGMA, MRTN-CT-2004-5652 
and  L.G. was supported by a grant of the ANR Program GIMP, ANR-05-BLAN-0029-01.

\end{document}